# Sub-picoliter Traceability of Microdroplet Gravimetry and Microscopy

Lindsay C. C. Elliott,[†,‡] Adam L. Pintar,[†] Craig R. Copeland,[†] Thomas B. Renegar,[†] Ronald G. Dixson,[†] B. Robert Ilic,[†] R. Michael Verkouteren,[†] and Samuel M. Stavis[*,†]

[†]National Institute of Standards and Technology, Gaithersburg, Maryland 20899, United States
[‡]University of Maryland, College Park, Maryland 20742, United States
[*]Email: samuel.stavis@nist.gov

**ABSTRACT:** Gravimetry typically lacks the resolution to measure single microdroplets, whereas microscopy is often inaccurate beyond the resolution limit. To address these issues, we advance and integrate these complementary methods, introducing simultaneous measurements of the same microdroplets, comprehensive calibrations that are independently traceable to the International System of Units (SI), and Monte-Carlo evaluations of volumetric uncertainty. We achieve sub-picoliter agreement of measurements of microdroplets in flight with volumes of approximately 70 pL, with ensemble gravimetry and optical microscopy both yielding 95% coverage intervals of ± 0.6 pL, or relative uncertainties of ± 0.9%, and root-mean-square deviations of mean values between the two methods of 0.2 pL or 0.3%. These uncertainties match previous gravimetry results and improve upon previous microscopy results by an order of magnitude. Gravimetry precision depends on the continuity of droplet formation, whereas microscopy accuracy requires that optical diffraction from an edge reference matches that from a microdroplet. Applying our microscopy method, we jet and image water microdroplets suspending fluorescent nanoplastics, count nanoplastic particles after deposition and evaporation, and transfer volumetric traceability to number concentration. We expect that our methods will impact diverse fields involving dimensional metrology and volumetric analysis of microdroplets, including inkjet microfabrication, disease transmission, and industrial sprays.

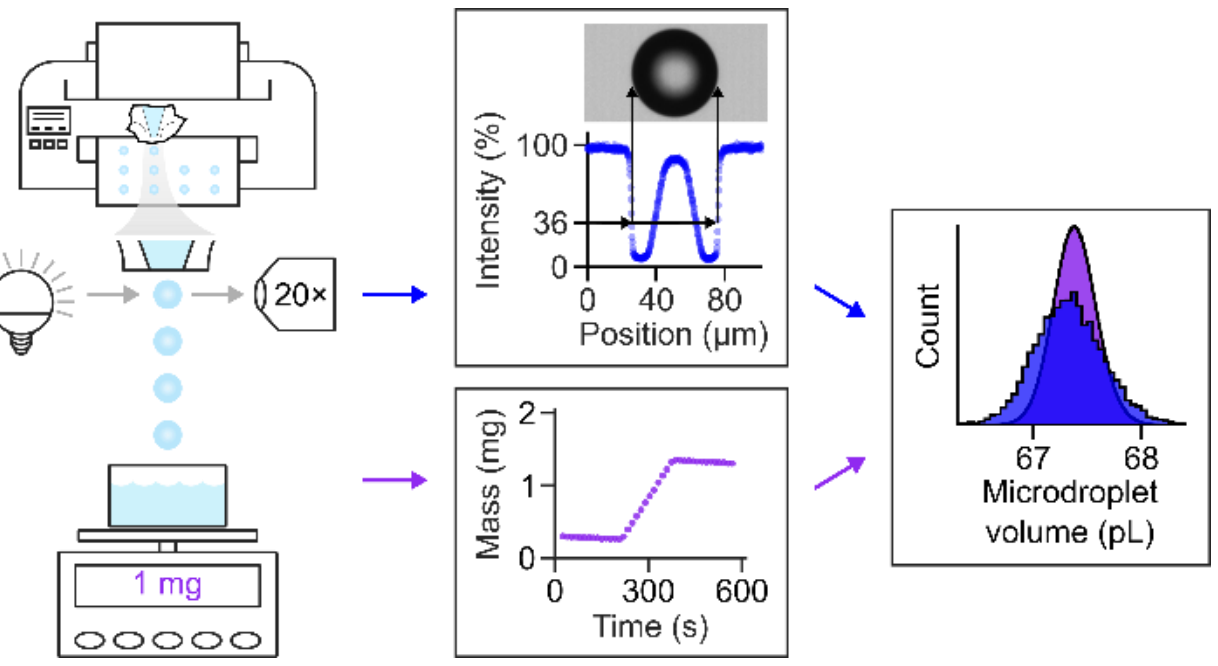

Measurements of microdroplets are important in many applications, including inkjet microfabrication,[1–3] disease transmission,[4–8] and industrial sprays.[9–12] Optical microscopy is a common method for imaging the dimensions and motion of single microdroplets, but current methods have two main limitations. First, microscopy measurements typically lack a direct connection to gravimetry, which defines the state of the art for ensemble measurements of microdroplet mass that are traceable to the International System of Units (SI).[13–15] Traceable measurements of density are then necessary to determine volume. Second, microscopy measurements of microdroplet dimensions typically lack a deep root in optics beyond the resolution limit corresponding to the imaging wavelength. To address these issues, we advance and integrate these methods. The key to improving microscopy toward parity with gravimetry is the accurate localization of microdroplet edges.[15,16]

Localization of edge position by optical measurement is the basis of dimensional metrology across many applications, including inkjet microfabrication.[9–11,17–25] Image edges have finite widths that depend primarily on the imaging wavelengths, numerical apertures, and object depths. As Fresnel theorized and Sommerfeld proved, the intensity distribution from diffraction at a knife edge, an approximation of an infinite half-plane, is due to the interference of light waves.[26,27] In the far field, where the Fraunhofer approximation is good, the position of a knife edge corresponds to a relative intensity of 50% for a spatially incoherent system and to a relative intensity of 25% for a spatially coherent system.[26,28] We refer to this relative intensity as the edge threshold. Models of diffraction at the edges of complex objects or in complex optical systems can require numerical solutions and can be difficult to apply to experimental systems.[29–33] Therefore, an accurate determination of the edge threshold, among other microscopy parameters, generally requires either an object and optical system that are highly ideal, or calibration of an experimental microscope using standards. In contrast, microscopy measurements of microdroplet dimensions have involved either arbitrary values of edge thresholds or incomplete characterization of reference objects. The resulting localization accuracy is questionable, leading to potential overconfidence in precise but inaccurate edge positions.

We solve this central problem within the scope of a comprehensive study. To test the accuracy of edge localization, we perform the first simultaneous measurements of the same microdroplets by ensemble gravimetry and optical microscopy. For the first time, each method is independently traceable to the SI through comprehensive calibrations and Monte-Carlo evaluations of uncertainty. We image microdroplets in flight by shadowgraphy, a transmission microscopy method of measuring microdroplet size, speed, and fluid properties.[17,34] To determine the edge threshold, we evaluate a double knife edge and dielectric microspheres as microscopy standards (Figure 1A), finding that matching the optical diffraction from the reference object to that of experimental microdroplets improves accuracy. In this way, we present a method that is practical to implement in microscope systems of arbitrary characteristics, including imaging wavelength, spatial coherence, field depth, and intrinsic aberrations. We develop our methods by inkjet printing microdroplets of cyclopentanol with a volume of approximately 70 pL. For both gravimetry and microscopy, we achieve 95% coverage intervals of ±0.6 pL, or relative uncertainties of ±0.9%, and root-mean-square deviations of mean values between the two methods of 0.2 pL, or 0.3% across several experiments. Our microscopy results improve upon previous microscopy results by an order of magnitude and are independent of gravimetry.[15] In an application of our microscopy method, we jet and image water microdroplets suspending fluorescent nanoplastics, count nanoplastic particles after deposition, and transfer volumetric traceability to the number concentrations of single microdroplets.

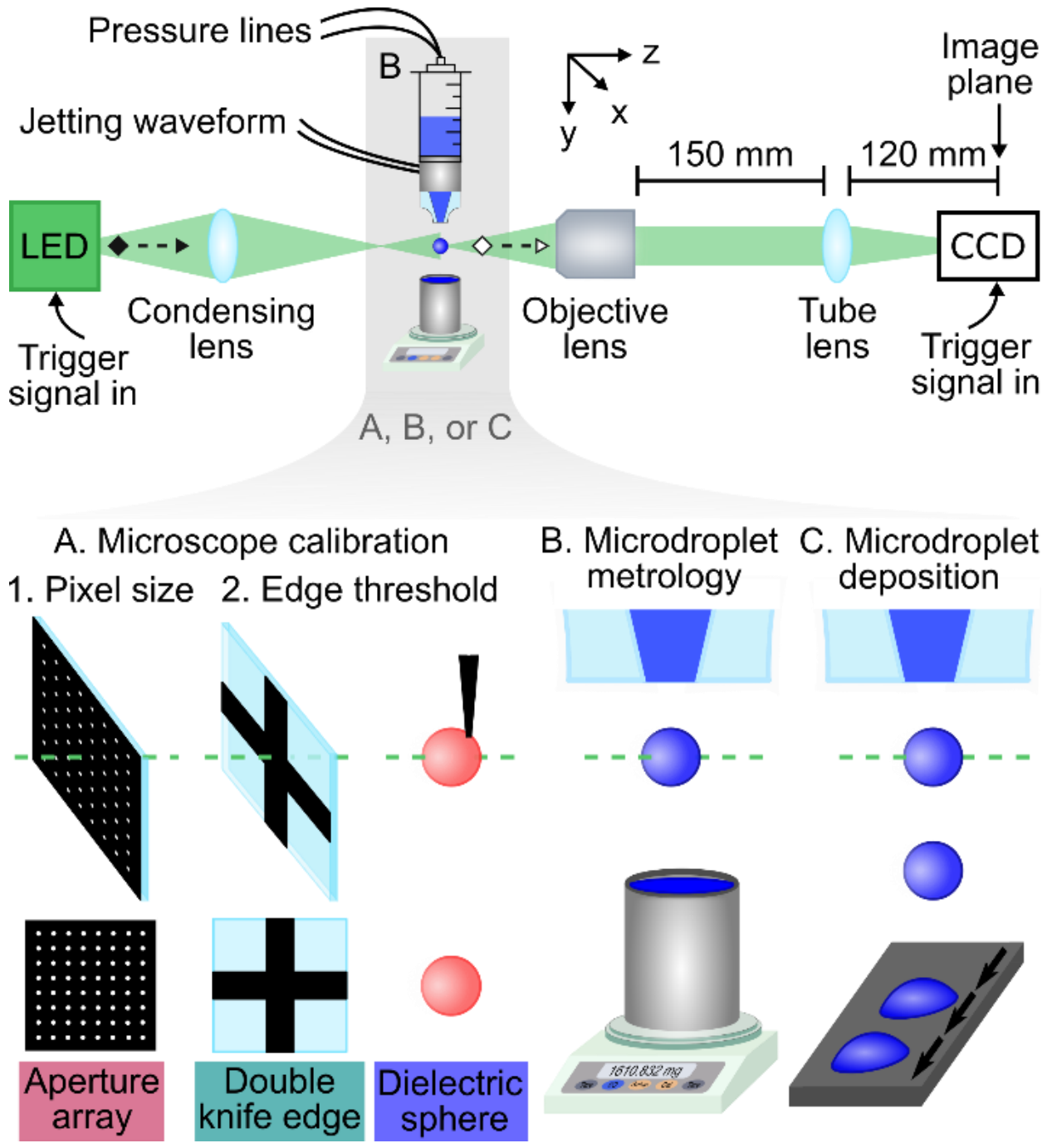

**Figure 1.** Experimental system (Table S1) integrating inkjet printing, optical microscopy, and ensemble gravimetry with modes of operation A1, A2, B, and C. The schematic depicts (gray box) mode B with a microdroplet in the optical path and balance underneath. The optical system has (black diamond) illumination and (white diamond) collection paths. Pressure and electronic systems supply pressure lines, a jetting waveform, and a trigger signal (Scheme S1). The object in the optical path can be either an aperture array, double knife edge, dielectric sphere, or microdroplet in flight. Standards for pixel size and edge threshold enable microscope calibration, a gravimeter with a weighing vessel allows microdroplet metrology, and a substrate on a translation stage implements microdroplet deposition. Color coding for labels of reference objects matches other figures and schemes.

## MATERIALS AND METHODS

**Overview.** An inkjet dispenser with a piezoelectric actuator forms microdroplets of cyclopentanol for microdroplet metrology or aqueous suspensions of fluorescent nanoplastics for microdroplet deposition. Cyclopentanol has an ideal viscosity for jetting, evaporates slowly, and has low hygroscopicity, whereas water is more challenging to jet and evaporates quickly but is relevant to more applications. The experimental system images single microdroplets of cyclopentanol using a shadowgraphy microscope and simultaneously weighs an accumulating ensemble of the same cyclopentanol microdroplets using a submicrogram balance (Figure 1B). The water microdroplets suspend fluorescent nanoplastics, consisting of polystyrene particles with nominal diameters of 200 or 500 nm, carrying fluorophores, and having surfaces with carboxylate functionalization (Supporting Information). The shadowgraphy microscope images water microdroplets before deposition into linear arrays (Figure 1C). After evaporation of the water microdroplets, a fluorescence microscope images nanoplastic particles in deposition regions (Supporting Information).

**Inkjet system.** We implement the jetting methods of previous studies[13–15] with some deviations, notably a reduction of microdroplet speed and a different dispenser assembly. The dispenser forms microdroplets at relatively low frequencies of less than 1 kHz by a backing pressure and a voltage waveform that result in stable jetting. A cartridge dispenser with a polypropylene syringe body contains the jetting medium and attaches directly to the dispenser (Figure 1B,C). We use different dispensers to jet cyclopentanol and aqueous suspensions of fluorescent nanoplastics, rinsing the latter dispenser copiously with pure water between the different suspensions. The inkjet pressure system provides backing and purge pressure to the dispenser (Scheme S1, Table S2). To jet microdroplets, the drive electronics controller produces a voltage waveform that actuates the piezoelectric dispenser (Scheme S1, Table S2). The jetting software controls waveform parameters and runs jetting scripts. The drive electronics controller triggers the linear stage and the microscope.

**Gravimetry.** The inkjet system dispenses microdroplets with a mean mass of approximately 65 ng onto a balance for ensemble gravimetry (Figure 1B). A vessel containing the jetting liquid rests on the weighing pan. The dispenser nozzle is within a few millimeters of the liquid level. The balance has sub-microgram resolution and performs instantaneous measurements of mass at a rate of approximately 1 Hz. A data collection script records the time and mass at a rate of approximately 0.1 Hz. An enclosure reduces noise from air currents while weighing cyclopentanol microdroplets. The balance rests on vibration dampers of viscoelastic polyurethane on a floating optical table and has an electrical ground through the table.

**Microscopy.** A shadowgraphy microscope records transmission micrographs of microdroplets with a mean diameter of approximately 50 $\mu$m in flight (Figure 1). In the illumination path, the source is a light-emitting diode (LED) that is partially coherent and has an array size of 2 by 2 mm, a peak emission wavelength of approximately 520 nm (Figure S1), and a power output of 1 W after collimation. A condensing lens collects and focuses the illumination slightly before the sample plane, where microdroplets travel through the optical path. The condensing lens position maximizes the mean intensity at the imaging sensor. In the collection path, an objective lens with air immersion, a nominal magnification of 20×, a numerical aperture of 0.22, and a working distance of 12 mm collects light. A matching tube lens forms an image on the sensor of a charge-coupled device (CCD) camera. The CCD array has 1600 by 1200 pixels, each with an on-chip size of 4.4 by 4.4 $\mu$m, an analog-to-digital output of eight bits, and a maximum imaging rate of 15 Hz. The 150 mm spacing between the objective lens and the tube lens, and the 120 mm spacing between the tube lens and the plane of the imaging sensor, meet the design specifications of the tube lens. In this way, the experimental magnification is approximately equal to the nominal magnification, and the imaging system implements some of the aberration corrections of the lens pair. The image pixel size is approximately five times smaller than the Abbe criterion, oversampling the feature edges with respect to the Nyquist criterion. The inkjet trigger synchronizes the jetting and imaging systems. The trigger divider in the jetting software sets the trigger frequency at 10 Hz, collecting images for 10% or 100% of inkjet waveforms and of microdroplets. The LED illumination time is 100 ms. The CCD exposure time is 3.938 $\mu$s, which is the fastest timescale of the imaging process.

**Microscopy standards**. An aperture array enables the calibration of magnification and distortion (Figure 1A1).[35] The apertures have nominal diameters of 0.5 $\mu$m in a bilayer of titanium with a thickness of 5 nm and platinum with a thickness of 80 nm, on a silica coverslip with a thickness of 0.17 mm (Table S1). The square array has a pitch of 5.001 45 $\mu$m ± 0.000 54 $\mu$m that is traceable to the SI (Table S3).[36] Two reference objects enable the calibration of edge thresholds and tests of field dependence (Figure 1A2). The first is a double knife edge, with an edge separation of approximately 25.0 $\mu$m, in a chromium film with a thickness of 100 nm, on a quartz substrate with a thickness of 2.3 mm (Table S1). We measure the edge separation (Table S3) by critical-dimension atomic-force microscopy that is traceable to the SI. The second reference object is one of four polystyrene microparticles from two populations with a mean diameter of 50.2 $\mu$m ± 0.3 $\mu$m and standard deviation of 0.5 $\mu$m, or with a mean diameter of 39.94 $\mu$m ± 0.35 $\mu$m and standard deviation of 0.52 $\mu$m (Table S1). We measure the dielectric spheres by optical profilometry, calibrating these measurements using step height standards that are traceable to the SI (Table S4).

**Microdroplet metrology.** A fresh dispenser yields stable jetting, with some deviations that we revisit, for simultaneous gravimetry and microscopy. The system jets at 100 Hz and images 10% of cyclopentanol microdroplets in flight at a delay of 500 $\mu$s after jetting (Table S5). The microdroplet speed is 0.2 m/s in experiments 1 through 4 and 0.4 m/s in experiments 5 and 6. The number of microdroplet images depends on the jetting mode and is 399 for burst mode and approximately 2200 for continuous mode, with analysis of every 5$^{th}$ or 10$^{th}$ image, respectively.

**Microdroplet deposition.** The system jets at 10 Hz and images 100% of water microdroplets suspending fluorescent nanoplastics in flight at a delay of 800 $\mu$s after jetting. The microdroplet speed is 0.3 m/s, which is between the two speeds for microdroplet metrology (Table S5). In each experiment, the dispenser prints a linear array of 20 microdroplets on a translating borosilicate substrate.

**Model fits.** We fit models to data using the method of damped least-squares with uniform weighting.

**Uncertainty evaluation.** We define a measurement process which maps input quantities to an output quantity.[37] Probability distributions describe uncertainties of the inputs, which propagate to uncertainty of the output by a Monte-Carlo method.[37] We report the final uncertainty of volume as a 95% coverage interval, corresponding to the 2.5% and 97.5% percentiles of the output distribution, and use this metric to calculate relative uncertainty. The probability distributions of the inputs and the number of Monte-Carlo trials depend on the details of four different measurements of microdroplet volume. We analyze sources of uncertainty and sensitivity of volume to errors (Supporting Information, Figure S2).

## RESULTS AND DISCUSSION

Microdroplet metrology (Figure 1B) and microdroplet deposition (Figure 1C) involve competing constraints between inkjet printing, gravimetry, and microscopy. Inkjet printing forms microdroplets of reproducible size and speed, with better reproducibility at speeds above 1 m/s. Gravimetry requires stability of jetting and evaporation, which is also better at higher speed. Microscopy requires the stability of microdroplet shape and position, which is better at higher speed but also suffers from motion blur, which is better at lower speed. Experimental parameters (Table S5) that balance these constraints result in microdroplet speeds ranging from 0.2 to 0.4 m/s in the *y* direction, from calculations of microdroplet positions at multiple times. The microdroplets travel between 0.8 and 1.6 $\mu$m during the exposure time, or between 4 and 8 pixels for a round image of 230 pixels, which corresponds to a microdroplet diameter of 50 $\mu$m. Control experiments to vary the exposure time and speed (Figure S3) show that motion blur at these speeds has little effect on diameter. Sweeping through the delay time between jetting and imaging enables collection of videos of microdroplet formation and oscillation, informing selection of the trigger delay for optimal measurement of spherical microdroplets, at a stable position, after damping of oscillation, and with minimal evaporation and background from the dispenser (Figure S4).

**Gravimetry.** *Overview.* Ensemble gravimetry involves analysis of mass as a function of time, featuring two main trends. Fundamentally, mass increases as a microdroplet ensemble accumulates during jetting. Simultaneously, mass decreases as liquid evaporates from the weighing vessel, requiring analysis of baseline slopes.

*Calibration.* We perform internal calibration and linearization routines on our balance and weigh reference masses (Table S6) to achieve traceability to the SI (Supporting Information).

*Operation.* We collect data in two jetting modes, requiring different analyses. The first is continuous mode, in which the main parameter is the frequency of jetting. The second is burst mode, in which the main parameter is the number of microdroplets. In continuous mode, the nominal frequency is 100 Hz, which an oscilloscope measures at 100.009 Hz with a standard deviation of 0.0051 Hz (Table S7), and jetting starts and stops by manual control. In burst mode, a script controls jetting in four bursts of 999 microdroplets (Supporting Information). We collect data only after observing a baseline trend with steady decreases of mass by 2 to 3 µg for an interval of 10 s (Figure 2a). To reduce uncertainty in fitting, at least 20 data points in each section of before, during, and after jetting, comprise data sets.

*Analysis.* We advance the gravimetry of an ensemble of microdroplets by our Monte-Carlo evaluation of uncertainty. To begin the analysis, application of the calibration function to the source mass yields mass with uncertainty (Supporting Information), defining a normal distribution for each data point. Sampling from this distribution yields mass data for analysis (Table S7, Scheme S2a). Correction of source time yields time without uncertainty in units of seconds (Supporting

Information, Scheme S2b). In both modes, fitting 20 different subsets of the 42 baseline data points accounts for the full range of possible rates of evaporation (Scheme S2c). Piecewise linear models fit mass as a function of time (Figure 2a, Scheme S2d). For this analysis, the necessary assumption is that the evaporation rate while jetting is a linear combination of the rates before and after jetting. Each subset of 22 data points varies the weighting of the before and after baseline slope (Scheme S2c). Then, in the analysis of continuous-mode data, a linear model fits the jetting period. Subtraction of the baseline slope from the jetting slope, and then division by the jetting frequency, yields the mean mass of the microdroplets (Scheme S2e). For burst-mode data, the baseline slope extrapolates the before and after baselines to the jetting midpoint. Division of the difference at the midpoint by the number of microdroplets yields mean mass (Scheme S2e). For both jetting modes, cyclopentanol density follows a uniform distribution (Figure 2b, Scheme S2f) from measurements that are traceable to the SI with a dependence on ambient temperature (Table S8) ranging from 19.5 to 22.5 °C (Table S7).[38] The uncertainty contribution from the range of ambient temperature dominates the total uncertainty for cyclopentanol density. Division of mean mass by a value from the density distribution (Scheme S2f) yields mean volume (Figure 2c). The higher uncertainty of volume for burst mode relative to continuous mode (Figure 2c) is due to the extrapolation, which contributes a relatively large component of uncertainty.

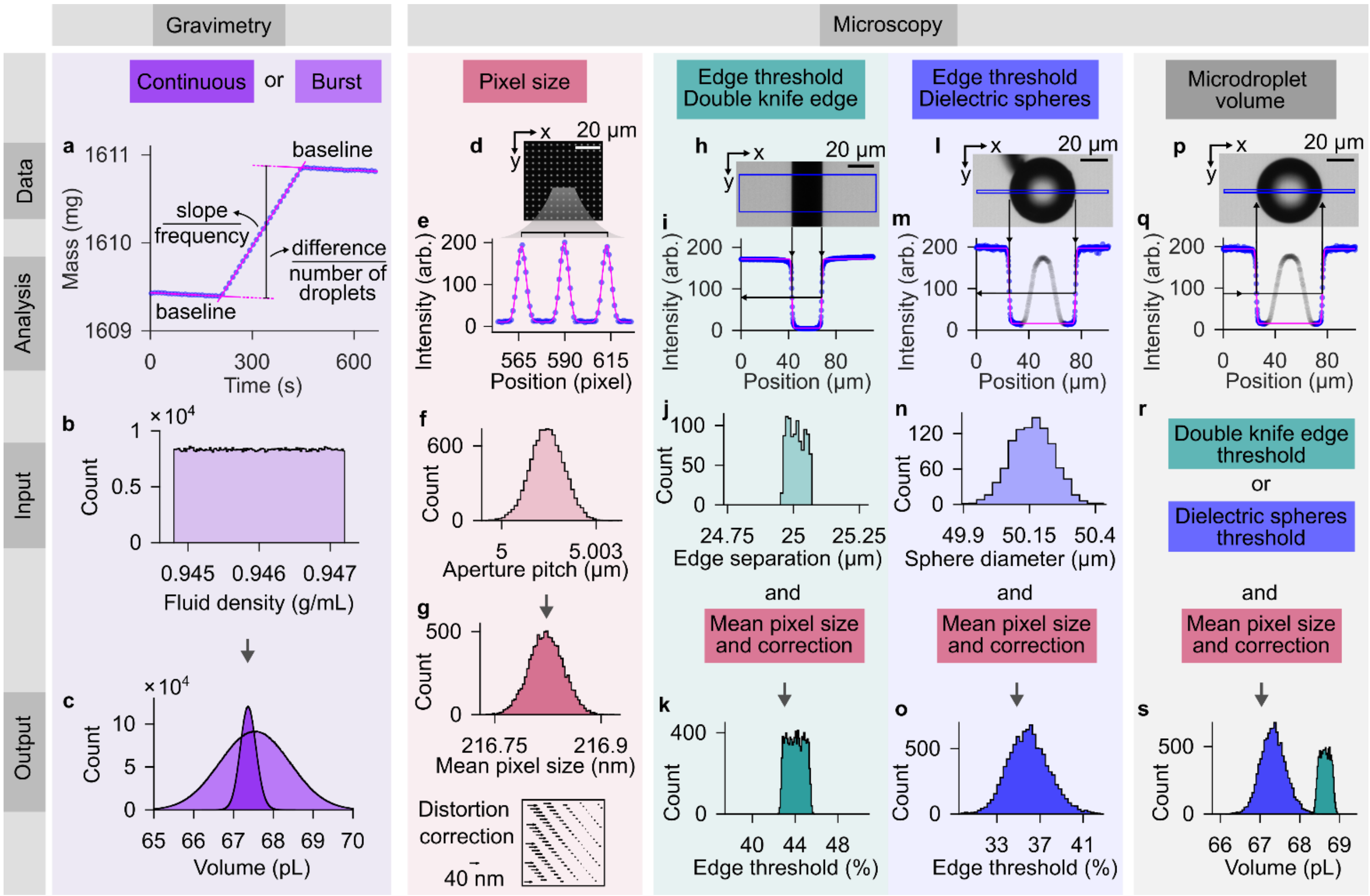

**Figure 2.** Measurement workflow with major features of data, analysis, input, and output for gravimetry and microscopy. Histograms depict the Monte-Carlo evaluation of uncertainty. Analysis plots include (blue circles) data in fits, (light gray circles) data not in fits, (solid magenta lines) fits, and (dot magenta lines) fit extrapolations. Color coding matches other figures. Horizontal axis scales match in (c,s,j,n,k,o). Distribution details are in Table S7. (a−c) Gravimetry. (a) Plot of mass data with linear fits and extrapolations. (b) Histogram of cyclopentanol density. (c) Histograms of volume from (dark purple) continuous and (light purple) burst jetting modes. (d−s) Microscopy. (d−g) Pixel size. (d) Transmission micrograph of aperture array. (e) Plot of intensity across three apertures with Gaussian fits. (f) Histogram of aperture pitch. (g) Histogram of mean pixel size and plot of distortion correction. (h−o) Edge threshold. (h,l) Transmission micrograph of double knife edge and dielectric sphere. A blue box outlines the pixel rows for analysis. (i,m) Plot of intensity across edge reference with a double error function fit. (j,n) Histogram of the reference dimension for the double knife edge and for a single dielectric sphere, and pixel size output. (k,o) Histogram of edge threshold. (p−s) Microdroplet volume. (p) Transmission micrograph of cyclopentanol microdroplets in flight. A blue box outlines the pixel rows around the equator for analysis. (q) Plot of intensity across the microdroplet with a double error function fit. (r) Edge threshold and pixel size outputs. (s) Histograms of microdroplet volume from the (teal) double knife edge and (blue) dielectric sphere. Systematic effects on volume measurements are key results of this study.

**Microscopy.** *Overview.* Beyond the edge threshold, calibrations of magnification and distortion are necessary for accurate localization (Figure 1A).[35] Imaging near best focus and accounting for field dependences are important for such calibrations.[35,39,40] We align our reference objects, optimize the imaging conditions for each object (Supporting Information), and image reference objects and microdroplets through focus in axial increments of 1 $\mu$m. Collection of each series of images through focus occurs in the same direction of travel and with 100 replicate images at each *z* position. Aperture images indicate best focus by the highest mean value of integrated brightness (Figure S5a). The images of edges indicate best focus by the narrowest value of width (Figure S6a). Pixel size (Figure S5b), edge threshold (Figure S6b), and apparent microdroplet volume with a constant edge threshold (Figure S6c) are all sensitive to *z* position. Therefore, achieving accuracy requires the selection of images near best focus. Measurements of field curvature, field dependence of edge width, and position dependence of edge threshold and volume (Figure S7) show small to negligible effects.

*Pixel size.* Analysis of the apparent positions of apertures, in combination with a distortion correction and traceable pitch,[35,36] yields an image pixel size. Asymmetric Gaussian functions approximate the aperture images and extract apparent positions. We depict this analysis by a transmission micrograph of an aperture array (Figure 2d) and by a line profile and fit of the images of three apertures (Figure 2e). The mean and uncertainty of the array pitch define a normal distribution (Figure 2f, Table S3), which yields the mean pixel size and distortion correction (Figure 2g). These quantities depend on the orientation in the $x$–$y$ plane, so we use the mean pixel size in the *x* direction to match the direction of the microdroplet equator, minimizing error from motion blur (Figure S3b). To construct a normal distribution around the mean pixel size, we select four *z* positions surrounding best focus with the aperture array, totaling 400 images (Figure S5b,c) and yielding a mean of 216.82 nm with a standard deviation of 0.02 nm. A single map suffices for distortion correction, as the variation between maps from different aperture images is negligible (Figure 2g, Table S7). Vectors point in the direction of the distortion correction, which is mostly inward and slightly offset from the center, showing a typical misalignment and field dependence of the microscope. The maximum correction is approximately 65 nm, which is significant in comparison to the mean pixel size. Neglect of this correction would yield a volume error of up to 0.4% for the microdroplets in our study.

*Image analysis.* Analysis of edge reference and microdroplet images near best focus allows calculations of edge threshold and microdroplet volume. We perform Monte-Carlo evaluations of uncertainty. Each trial involves different light and dark background images, intensity profile data, mean pixel size, and edge threshold for microdroplets. The analysis of edge threshold uses a different source image for each trial to capture variation from image to image. For each source image, edge reference, and microdroplet, the uncertainty evaluations involve $10^3$ trials with the double knife edge threshold, or $10^4$ trials with the dielectric sphere threshold and microdroplet volume. In analyses of round images with only $10^3$ trials, larger variation shows that $10^4$ trials are necessary to reproduce the edge threshold distribution for dielectric sphere images and volume distribution for microdroplet images. The larger variation for dielectric spheres and microdroplets is due to the necessary use of fewer rows of pixels than with the double knife edge, 6 in comparison to 177, in analysis of the intensity profile data. A height of 6 pixel rows around the *y* position of the equator balances sampling a small enough number of pixel rows to avoid a significant systematic underestimate of diameter, with sampling a large enough number of pixel rows to reduce components of uncertainty from random effects, particularly for edge width (Figure S8).

To start the analyses of edge threshold and microdroplet images, the mean and standard deviation of pixel values from $10^3$ images without an object in the field of view, and without or with constant illumination, forms dark and bright backgrounds, respectively, and quantifies the variability (Scheme S3a; Table S7). For each pixel, subtraction of the dark background from the source image, and normalization by the bright background, processes the image for subsequent analysis (Scheme S3b; Figure 2h,l,p). A region of interest (blue boxes in Figure 2h,l,p) defines an area for reducing image data to an intensity profile, by averaging columns of pixels to obtain a mean and standard error (Scheme S3c). For the double knife edge, the region has a width of 406 pixels, a height of 177 pixels, and is concentric with both the microdroplet images and the images from critical-dimension atomic-force microscopy. The column mean and standard error define a normal distribution. For dielectric spheres and microdroplets, the region has a width of at least 580 pixels, a height of 6 pixels around the equator, and is concentric with each feature. The column mean and standard error define a Student *t* distribution. The mean pixel size and distortion correction convert the position vector from pixel to SI units (Scheme S3d). Double error functions approximate the intensity profiles, excluding the central region of high intensity from microlensing (Scheme S3e; Figure 2i,m,q), to extract parameters and determine the edge threshold and microdroplet volume.

*Edge thresholds.* The combination of edge reference dimensions and intensity profile fits yields edge thresholds. The reference dimension from either edge separation or sphere diameter (Figure 2j,n; Table S4) corresponds to a threshold intensity (Scheme S3f). The minimum and maximum values from the double error function convert the threshold intensity value to a percentage (Scheme S3g). The threshold depends on the orientation in the $x$–$y$ plane, so we analyze the edge threshold only in the *x* direction to match the corresponding orientation of the microdroplet analysis. Differences between the uncertainties of the reference dimensions (Figure 2j,n) and between the properties of the edge references are evident in the resulting distributions of edge threshold (Figure 2k,o). The double knife edge yields an edge threshold mean of 44.04% and a standard deviation of 0.80%, whereas the dielectric spheres yield an edge threshold mean of 35.94% and a standard deviation of 1.91%. These key results show the potential for a systematic error due to a mismatching edge reference. The subsequent comparison of gravimetry and microscopy enables a deeper understanding of this critical issue.

*Microdroplet volume.* Either edge threshold allows calculation of the microdroplet volume (Figure 2r). Minimum and maximum values from the fit of the double error function to the microdroplet profile convert edge threshold to an intensity value. The $x$ positions corresponding to this intensity value yield the microdroplet diameter (Scheme S3f). Other edges of the image have more motion blur, so we use the diameter only at the equator to calculate volume, assuming sphericity (Scheme S3g; Figure 2s). With faster imaging, all lateral edges of a microdroplet can yield measurements of size and shape, assuming axisymmetry.[17] The volume of a sphere cubes the diameter, tripling its relative uncertainty. For example, in metrology experiment 1 (Table S9), a diameter of 50.53 $\mu$m with a 95% coverage interval of ±0.15 $\mu$m corresponds to a volume of 67.56 pL with a 95% coverage interval of ±0.63 pL for relative uncertainties of ±0.30 and ±0.93%. Nonetheless, the relative uncertainty of volume for microscopy is similar to that of gravimetry.

**Microdroplet Metrology.** Our experimental system enables simultaneous gravimetry and microscopy of the volume of cyclopentanol microdroplets in six experiments, three in continuous mode and three in burst mode (Figure 3). We compare results from gravimetry and microscopy (Figure 3a, Table S9), which are fundamentally different measurements. Gravimetry yields the mean volume from an ensemble of microdroplets, whereas microscopy yields images of single microdroplets. The microscopy data in chronological order show significant changes in the jetting process.

Among experiments 1 through 4, the mean volume varies by up to 0.3 pL or 0.4% with data collection spanning 1 h (Table S10), which is comparable to the differences between the results from gravimetry and from microscopy with dielectric spheres within each experiment. This variation can result from a changing dispenser surface, emphasizing the need for simultaneous measurements by the two methods for direct comparison. Comparing gravimetry results within jetting modes, larger uncertainties of mean volume in experiments 2 and 5, and experiments 1 and 4 to a lesser extent, are due primarily to the evaporation rate, as different rates before and after jetting can result in bimodal distributions of volume (Figure S9b). For the edge threshold from dielectric spheres, gravimetry and microscopy show a root-mean-square deviation of 0.31% or 0.21 pL (Table S9), which is a fraction of the width of the 95% coverage intervals. This demonstrates a near equality of the mean microdroplet volume in gravimetry and single microdroplet volume in microscopy.

The results of experiment 3 emphasize an important distinction between volumetric analysis of single microdroplets and of a microdroplet ensemble. The microscopy results show an abrupt decrease of the volume of single microdroplets during the experiment (Figure 3b), indicating that jetting temporarily ceases and then resumes to form smaller microdroplets. This transient event may be due to a change of dispenser wetting. In contrast, for gravimetry, analysis of partial segments of the jetting period is unable to extract the two different volumes, showing only a larger uncertainty due to the smaller sets of data.

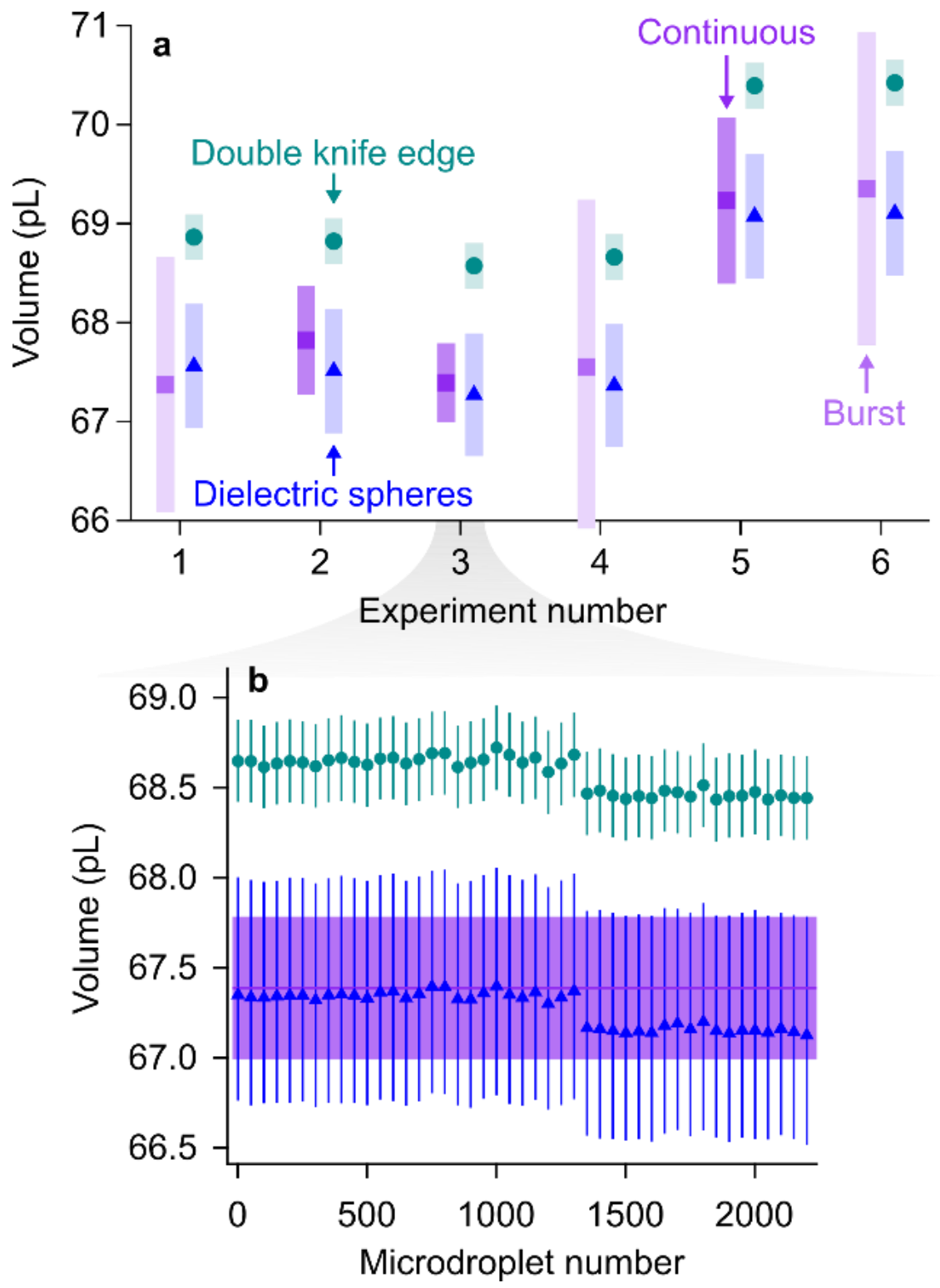

**Figure 3.** Microdroplet metrology results from simultaneous gravimetry and microscopy measurements. (a) Plot of microdroplet volume from all experiments including (dark purple square) continuous gravimetry; (light purple square) burst gravimetry; (teal circle) microscopy with double knife edge; and (blue triangle) microscopy with dielectric spheres. Mean and 95% coverage intervals are from distributions of mean volume from ensembles for gravimetry data and from pooled distributions of volume from single microdroplets, 224 in continuous mode and 75 in burst mode, for microscopy data. (b) Plot of experiment 3 with (purple line and shading) a mean volume from continuous gravimetry; single microdroplet volumes from microscopy with (teal circle) double knife edge and (blue triangle) dielectric spheres analyses, all with 95% coverage intervals. Details of each experiment are in Figure S9a.

*Edge diffraction.* The comparison of gravimetry and microscopy results allows further interpretation of the edge thresholds. A knife edge is the natural starting point for theoretical studies of diffraction[26,27] and has experimental advantages. In comparison to single microparticles, a double knife edge is easier to handle and measure (Supporting Information) and results in a lower uncertainty of the edge threshold. However, dielectric spheres yield a better agreement between our gravimetry and microscopy results, showing the importance of matching the properties of the edge reference with those of the microdroplet.

Edge properties, including conductivity, geometry, and volume effects due to imaging an object that is thicker than the depth of field, affect diffraction intensity and pattern, depending on the configuration, beam shape, and polarization of the illumination.[20,41–49] The remarkable work of Barnett and Harris[28] provides a useful context for interpreting our results. With conducting, absorbing, and transparent edges under unpolarized illumination, Barnett and Harris[28] observed that edge threshold is invariant to refractive index. Our illumination is unpolarized, so we expect a similar invariance. However, our microscope has a finite depth of field, in contrast to their optical system, so we expect our edge threshold to be sensitive to volume effects.

In this context, a polystyrene microparticle is optically similar to a cyclopentanol or water microdroplet, in that these objects are all dielectric spheres and transparent lenses, with refractive indices ranging from 1.33 to 1.59, and have similar diameters that are greater than the depth of field of our microscope. In contrast, the double knife edge is optically dissimilar to a microdroplet, being edges of a metal film with a thickness that is less than the depth of field of our microscope. The advantages of each type of object may be accessible by the fabrication of reference objects of an intermediate type.

**Number Concentrations of Single Microdroplets.** In an application of our microscopy method, we jet, image, and deposit water microdroplets suspending fluorescent nanoplastics. We perform three experiments, one with a particle diameter of 200 nm and two with a particle diameter of 500 nm. After deposition on a borosilicate substrate, the sessile microdroplets evaporate, leaving distinct regions of fluorescent nanoplastics within the bounding contact lines (Figure 4). The first few microdroplets deposit in proximity, due to a delay of linear stage motion after triggering (Figure S10a), and the first deposition regions have areal densities of single particles that are too high to count with low uncertainty (Figure S10b). We analyze the remaining 9 or 10 deposition regions for each experiment (Figure S10c). The volumes of single microdroplets in flight and the corresponding counts of nanoplastic particles in each deposition region yield number concentrations (Figure 4a,b). Counting single particles contributes a negligible uncertainty under experimental conditions of distinct deposition regions, a dilute suspension with low agglomeration, and a high signal-to-noise ratio (Supporting Information). In this way, we measure concentrations ranging from $1 \times 10^8$ to $5 \times 10^9$ particles/mL with 1% relative uncertainty (Figure 4c-e, Table S11), where the main uncertainty is from microdroplet volume.

The particle number concentrations of single microdroplet suspensions fluctuate and deviate from the nominal values of the bulk suspensions. We interpret these trends as being due to both microdroplet sampling of the bulk suspensions and two effects of jetting. First, particles adsorb to the surfaces of the jetting system, decreasing concentration. Second, water evaporates from the dispenser tip, increasing the local concentration. The concentration of particles with a diameter of 200 nm fluctuates around a constant value, suggesting a lesser effect of evaporation in this experiment, due to a shorter idle time before jetting. Both experiments with a particle diameter of 500 nm show decreasing concentrations after starting at up to an order of magnitude higher than that of the bulk suspension, indicating greater effects of evaporation during a longer idle time prior to jetting. The second experiment with a particle diameter of 500 nm approaches a concentration near that of the bulk suspension, whereas the concentrations from the experiment with a particle diameter of 200 nm are several times lower than that of the bulk suspension, possibly due to greater adsorption of smaller particles than larger particles.

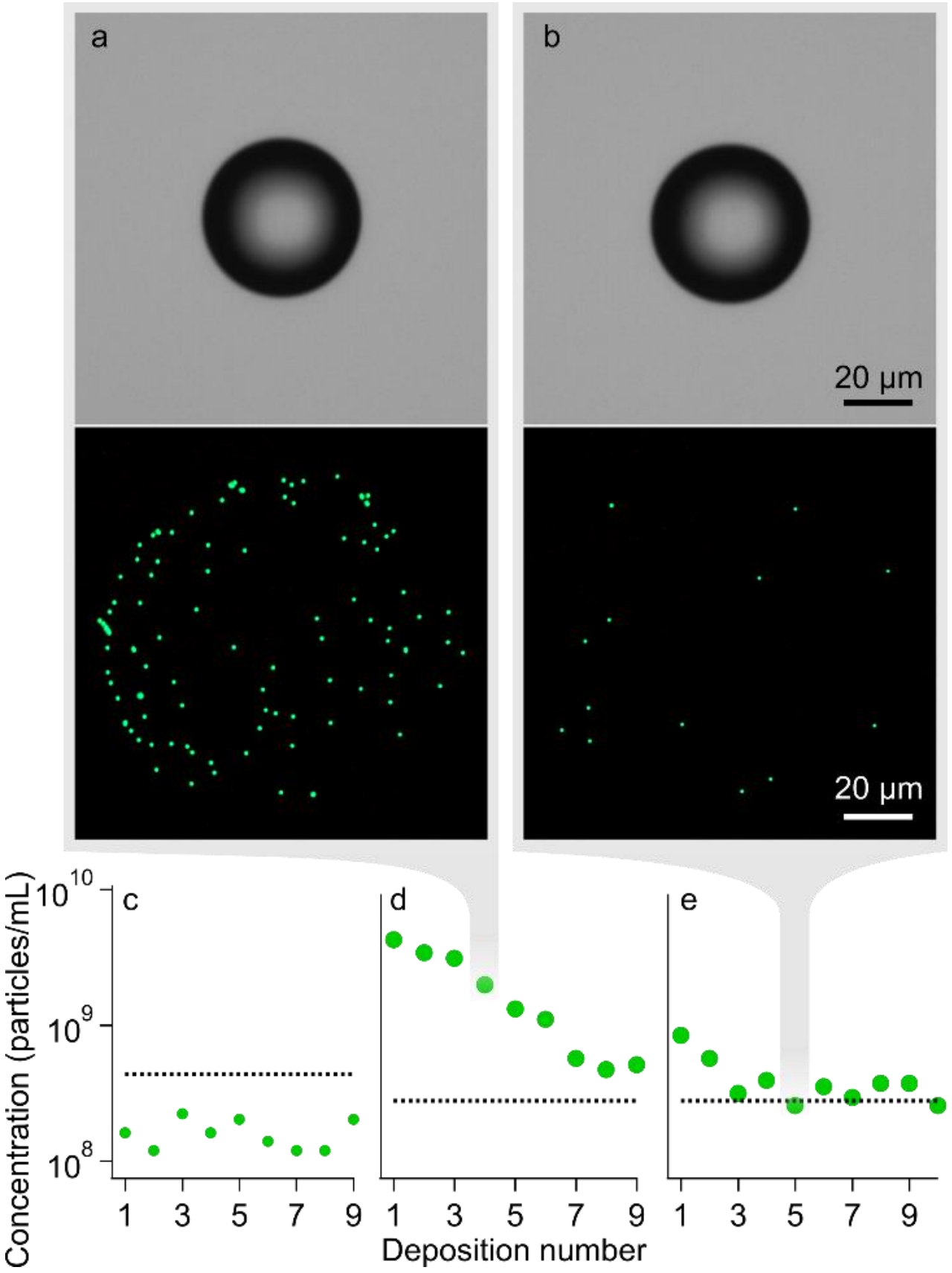

**Figure 4.** Number concentrations of single microdroplets. (a,b) Images from two depositions including (top row) transmission micrographs of water microdroplets in flight and (middle row) fluorescence micrographs in true color of nanoplastic particles on a borosilicate substrate. (c−e) Concentration plots for experiments with particle diameters of (small green circles) 200 nm and (large green circles) 500 nm. Dot lines are nominal values of $4.5 \times 10^8$ and $3 \times 10^8$ particles/mL for particle diameters of 200 and 500 nm, respectively, in bulk suspensions. 95% coverage intervals are smaller than the data markers.

## CONCLUSIONS

Ensemble gravimetry and optical microscopy are fundamentally different methods to measure the volume of microdroplets. Gravimetry has a short chain of traceability to the SI from mass through density to volume, and previous studies have identified the main sources of error. Gravimetry works best with the reproducible production of thousands of microdroplets, which is possible by jetting but yields a measure of mean mass that obscures any heterogeneity of the microdroplet population. In contrast, microscopy has the advantage of being able to measure the dimensions and motion of single microdroplets, which is useful in diverse applications. Moreover, recent advances in image analysis have overcome the resolution limit in many experimental systems. However, sources of error beyond the resolution limit remain under investigation. Therefore, our direct comparison of these two independent measurements yields complementary information benefitting both methods. Standards are critical to achieve accuracy and traceability in both measurements. Reference masses for gravimetry have existed since the earliest days of weights and measures, but suitable reference objects and corresponding calibrations for microscopy beyond the resolution limit have emerged only recently. Our improvement of microscopy accuracy to achieve sub-picoliter agreement with gravimetry indicates rapid progress, with the two measurements of microdroplet volumes of around 70 pL yielding 95% coverage intervals of ±0.6 pL and root-mean-square deviations of a mean volume of 0.2 pL or 0.3% between the two methods. Importantly, an edge threshold error would yield a larger relative effect for smaller microdroplets. Fabrication of knife edges and aperture arrays from the top down, and synthesis of dielectric spheres from the bottom up, are complementary approaches to form standards for the calibration of optical microscopes. The fabrication of curving and complex microstructures in dielectric materials presents a future opportunity to form new reference objects.[50–52] Building on this capability, future studies could elucidate imaging effects from microscale objects with different shapes and sizes with respect to the depth of field, ranging from microdroplets to biological cells.

Our methods have many potential uses. Applying our inkjet printing capability, we microdeposit arrays of water microdroplets suspending fluorescent nanoplastics. After sessile microdroplet evaporation, we count nanoplastic particles and transfer volumetric traceability to the number concentrations of microdroplet suspensions. These values can deviate from that of a bulk suspension due to random sampling fluctuations and systematic jetting effects, with interesting implications. For example, our methods could directly apply to measurements of the viscosity of oscillating microdroplets suspending certain numbers of sub-micrometer or smaller particles.[34] Using our methods, and an imaging system that is 1 to 2 orders of magnitude faster than the present one, it would be possible to accurately measure the volume and speed of respiratory microdroplets that can transmit disease.[4,5] Finally, calibration of the field dependences of microscopy parameters would enable widefield imaging of a spray or cloud of microdroplets for studies in epidemiology, agriculture, and the environment.

## AUTHOR INFORMATION

### Contributions

S.M.S. supervised the study. L.C.C.E., R.M.V., and S.M.S. conceived the study. L.C.C.E. designed and implemented the experimental apparatus, with contributions from R.M.V., T.B.R., and S.M.S. A.L.P., L.C.C.E., C.R.C., and S.M.S. designed and implemented the data analysis. R.G.D. measured the double knife edge. B.R.I. fabricated the aperture array. C.R.C. measured the aperture array. T.B.R. calibrated the step height standards. L.C.C.E. and S.M.S. wrote the manuscript with contributions from A.L.P., C.R.C., R.G.D., and R.M.V.

### Notes

The authors declare no competing financial interest.

## ACKNOWLEDGMENTS

Paul DeRose, Natalia Farkas, Andrei Kolmakov, Thomas LeBrun, Mark McLean, Antonio Possolo, Dean Ripple, Keana Scott, Eric Shirley, Stephan Stranick, and Elizabeth Strychalski provided helpful comments. Sherry Sheckels and John Wright measured cyclopentanol density as a NIST measurement service. L.C.C.E. acknowledges support under the Cooperative Research Agreement between the University of Maryland and the NIST Center for Nanoscale Science and Technology, award number 70NANB14H209.

# Supporting Information

## Sub-picoliter Traceability of Microdroplet Gravimetry and Microscopy

Lindsay C. C. Elliott,[†,‡] Adam L. Pintar,[†] Craig R. Copeland,[†] Thomas B. Renegar,[†] Ronald G. Dixson,[†] B. Robert Ilic,[†] R. Michael Verkouteren,[†] and Samuel M. Stavis[*,†]

[†]National Institute of Standards and Technology, Gaithersburg, Maryland 20899, United States
[‡]University of Maryland, College Park, Maryland 20742, United States
[*]Email: samuel.stavis@nist.gov.

**CONTENTS**

**Uncertainty and sensitivity analysis**

*Overview.* Gravimetry and microscopy each have sources of uncertainty from random and systematic effects that are unique to the different methods of collecting and processing data. To understand these effects, and to inform the design of future experiments, we analyze the relative magnitude of these effects. We do this by comparing the uncertainty of microdroplet volume considering all sources, to the uncertainty of microdroplet volume with successive removal of each source, one at a time. To quantify possible systematic effects, such as incorrect values of jetting liquid density or edge threshold, we carry out a sensitivity analysis with successive variation of mean input values over relevant ranges.

*Gravimetry.* The analysis of uncertainty for both modes of gravimetry shows that most of the volume uncertainty results from mass uncertainty and almost none of the volume uncertainty results from the density uncertainty of the jetting liquid (Figure S2a). The total uncertainty of mass includes several components such as noise in the mass reading, balance reproducibility, and mass standard uncertainties. Noise in the mass reading has a nearly one-to-one positive correlation with volume uncertainty (Figure S2b), indicating the importance of reducing currents of air and stabilizing the rate of evaporation. Volume uncertainty increases as the number of data points in the linear fits decreases (Figure S2c), showing how gravimetry is a problematic measure of changes in microdroplet volume within an experiment. For continuous mode, fits of partial jetting periods have inherently larger uncertainty than fits of the total jetting period. For burst mode, it is impossible to investigate any changes in jetting because the analysis method compares the masses before and after jetting. For both modes, fits of fewer data points in the before and after baselines also result in increasing uncertainty of the measurement of microdroplet volume. Erroneous number of microdroplets, density, time, and jetting frequency in continuous mode all have the same nearly one-to-one negative correlation with volume error (Figure S2d).

On the basis of these results, the best way to reduce the uncertainty of a droplet volume measurement using dynamic ensemble gravimetry is to collect as many data points as possible for either mode, and to have a long duration of jetting in continuous mode, using a weighing vessel with ample capacity. Because of the extrapolation in the analysis of burst mode, collecting data points at a higher frequency rather than for longer periods is especially beneficial. It is possible to reduce the uncertainty of a mass measurement by calibrating a sub-microgram balance using mass standards with lower uncertainties, by minimizing noise of the reading by improving isolation from ambient vibration and by reducing exposure to air currents, and by stabilizing the evaporation rate.[1] Volumetric accuracy relies on independently measuring the jetting frequency in continuous mode, assuring the jetting of the correct number of microdroplets in burst mode, and, for both modes, measuring the temperature of the jetting liquid and the data collection rate.

*Microscopy.* The analysis of uncertainty for microscopy shows that most of the volume uncertainty is from edge threshold uncertainty and very little is from pixel size, intensity profile, and background uncertainties (Figure S2e, left). Most of the edge threshold uncertainty results from dimensional uncertainty of the edge reference and very little results from pixel size, intensity profile, and background uncertainties (Figure S2e, right). As a result, most of the volume uncertainty is due to dimensional uncertainty of the edge reference.

Sensitivity analysis shows that volume error correlates positively with edge threshold (Figure S2f) and reference dimension error (Figure S2g), with a maximum loss of volume of 15% from each source. The double sensitivity to percent changes in the double knife edge is because the double knife edge dimension is approximately half that of the dielectric spheres and the microdroplets. Volume error is sensitive to mean pixel size only for use of a constant value of edge threshold (Figure S2h), such as from an arbitrary or algorithmic definition. Interestingly, volume error is insensitive to errors of mean pixel size in combination with a varying edge threshold from a calculation with the same error of mean pixel size (Figure S2i). This cancellation of errors could be useful in experimental design. If calibration of the edge threshold relies on a reference object that matches the microdroplet in size and optical properties, such as a dielectric sphere, then a measurement of microdroplet volume can be accurate, even with an error of mean pixel size of up to 8%. However, we do not rely on this cancellation of errors, as we use such a reference object and an independent reference object, and we calibrate mean pixel size.

Finally, edge threshold has a nearly symmetric but opposite sensitivity to error in pixel size (Figure S2j) and to error in the dimension of the edge reference (Figure S2k), effects that correspond to errors of pixel size of 10 nm and edge reference dimension of -2 μm. We reject unreliable values for volume and threshold (light portions of lines in Figure S2f-k), which are either less than 5% and have large residuals in the fit with the double error function, or are greater than 50% and in conflict with optical theory. Violations of these upper and lower limits could result from erroneous values of edge threshold dimension or mean pixel size.

On the basis of these results, the best way to improve uncertainty of volume in microscopy is to use an edge reference with a low uncertainty, through some combination of fabrication specification and standard characterization. The best way to improve volumetric accuracy is to calibrate the edge threshold and to use an edge reference similar to the microdroplet in optical properties and dimensions.

*Errors.* Gravimetry and microscopy have different sources of uncertainty from random and systematic effects, so we expect that the total uncertainty will vary between the two measurements of microdroplet volume. However, if these two independent measurements of the same microdroplets are both accurate, as is our goal, then we expect to arrive at similar mean values of volume, with root-mean-square deviations that are consistent with the corresponding coverage intervals. Still, there are some systematic effects which we can identify and approximate, but for which we lack enough information to quantitatively correct in the evaluation of uncertainty, as is the case in nearly every measurement. To this end, we identify and estimate errors from four effects.

First, satellite microdroplets can accompany principal microdroplets,[2] and would increase ensemble mass by gravimetry in comparison to the volume of single microdroplets by microscopy. However, no satellites are visible in any of the images in the microdroplet metrology. Nanodroplets can have the same effect as satellites, although of smaller magnitude, and are visible in some images in the microdroplet metrology. On the basis of estimates of prevalence and volume, spurious nanodroplets increase mass only from 0.001 to 0.003%.

Second, initial microdroplets from jetting can be larger than subsequent microdroplets, known as the first droplets incongruity.[1] In burst mode with 999 microdroplets per burst, an arbitrary number due to a software limit, the effect can be significant, by one estimate increasing mass by 0.25%.[1] In contrast, in continuous mode with over 22000 microdroplets the effect is negligible and is remediable by exclusion of the first jetting data point from analysis. In cases where this effect is significant, jetting conditions can minimize the first droplets incongruity.

Third, missing microdroplets can depress mass in gravimetry in Experiment 3 from 0.05 to 0.13%.

Fourth, evaporation can decrease mass in gravimetry, even as the microdroplet videos show no signs of a significant loss of volume. Using the evaporation rate from the reservoir as a rough measure and adjusting for relative surface area, evaporation could decrease mass in gravimetry by approximately 0.02%.

Other potential errors are more difficult to estimate in our system but can have a significant effect on the results. These include inaccurate volume from gravimetry due to variations in evaporation rate or inaccurate thermometry, inaccurate volume from microscopy due to motion blur or microdroplet asphericity, or from any of the sources of volume error in the analysis of sensitivity (Figure S2d, S1f-i).

**Aperture array characterization**

We characterize the pitch of an aperture array by critical-dimension localization microscopy.[3] Our methods are similar to those of our previous study,[4] with the exception of the omission of camera calibration, which has a negligible effect at the experimental signal intensities. The aperture array is a working standard. We calibrate its pitch in reference to a master standard, which is another aperture array with a traceable pitch and uncertainty that we characterized previously by critical-dimension atomic-force microscopy.[5] We position the two aperture arrays side by side on an optical microscope stage for a series of replicate measurements under experimental conditions that are effectively identical.[6] We measure the working standard through focus and then we measure the master standard through focus, in the same lateral region of the optical imaging field to within a lateral misalignment of a few micrometers.[7] We repeat this measurement six times, laterally scanning the entire region of interest of the working standard for microdroplet imaging of approximately 100 by 180 μm through the same lateral region of the optical imaging field, obtaining images through focus at each lateral position. The six subregions of the overall region of interest of the working standard are independent. From each series of images of the aperture arrays through focus, we identify the axial position of average best focus, and localize the aperture images. We apply a similarity transformation of the working aperture positions to the master aperture positions to determine the relative scale between the two, yielding a pitch of the working aperture array of 5.001 45 μm ± 0.000 54 μm. This uncertainty is a 68% coverage interval of an approximately normal distribution. The uncertainty of scale factor in critical-dimension atomic-force microscopy dominates the total uncertainty of pitch, whereas sources of uncertainty from random effects such as lateral misalignment and environmental fluctuations are negligible.

**Double knife edge characterization**

We characterize the double knife edge using critical-dimension atomic-force microscopy. Just before measurement, we clean the double knife edge with piranha solution. This standard must fit into the imaging field of a single atomic-force micrograph, which is 40 by 40 μm, so we select a line feature with a width of approximately 25.0 μm (Figure 2h). The chromium pattern is opaque, and the quartz substrate is transparent. We follow the methods of previous studies[8,9] to collect and calibrate data that is traceable to the International System of Units (SI) through the National Institute of Standards and Technology (NIST), and to evaluate uncertainty. The two knife edges are at the two sides of the chromium feature, with sidewalls that are approximately trapezoidal, at angles of approximately 1.2 rad or 70° with respect to the surface of the quartz substrate (Figure S11). As a result of this edge geometry, we use values of width and edge roughness that depend on height to characterize the edge separation, selecting the maximum range to include all possible widths, with the lower and upper limit widths at 100% and 0% of the sidewall height, respectively (Table S3). This range is potentially decreasable by measurements of transmission as a function of film thickness, which would reduce the resulting uncertainty of microdroplet volume but would not improve the accuracy of the result from the double knife edge.

**Dielectric sphere characterization**

The particle size distribution of our polystyrene spheres is too large for us to draw a single particle from the population and simply assign a diameter with a corresponding uncertainty that is tolerably low for our purpose. Therefore, we further measure single particles using optical profilometry, or white-light interferometric microscopy. We measure the diameter of a spherical microparticle as the maximum difference in height between its apex and the underlying substrate. We presently describe the details of this measurement, which is non-trivial.

To prepare a single particle for this measurement, we evaporate a few microliters of the microparticle suspension on a silica substrate, reversibly adsorbing microparticles to the silica surface. We place the silica substrate on a stage in our imaging system and secure a tungsten microfilament in a syringe with a fixture to the cartridge holder, directly above the substrate. While visualizing the microfilament and microparticles with the imaging system, we use manual controls of stage position to push the microparticles against the microfilament tip until a clean microparticle desorbs from the silica substrate and adsorbs to the microfilament tip.

To measure edge threshold, we move the silica substrate out of the imaging field and position the microparticle at the center of the field (Figure 2l). We image the microparticle through focus at three orientations about the $y$ axis, each with a clear view of the microparticle diameter at the equator and with rotations of approximately $\pi/3$ rad or 60°. We find negligible variation in diameter between the three orientations. In addition, we verify that the size and shape of the polystyrene microparticles remain constant over time and upon adsorption to a silicon surface or tungsten microfilament (not shown).

To perform interferometric microscopy, we deposit the single microparticle onto the clean surface of a silicon substrate and image the microparticle. To calibrate the interferometric microscope, we image five standards for step height surrounding the nominal diameter of the microparticle of 40 or 50 μm. We generate a calibration curve[10] with the reference step heights and uncertainties from the calibration certificates and the source height data for each standard (Table S4).

To image each object, a microparticle or standard for step height, we collect and analyze height maps using commercial software for surface profilometry. Depending on the lateral extent of the object, we collect a single image or a composite image of up to 13 fields of view for some of the step height standards. We level each height map using either a three-point or least-squares optimization method. To extract the mean step height of each standard, we average from 49 to 1023 profiles, depending on the size of the standard, and then calculate the step height of the average profile.

To extract the mean diameter of a microparticle, we follow a similar procedure. However, as there are only a few data points around the apex of the sphere, we image the microparticle 10 times with small differences in position, calculate mean diameter from 9 data points near the apex in each scan, and average the 10 scan results. We use the calibration curve for step height to adjust this mean value and to calculate a standard deviation (Table S3). The resulting standard deviation of height is 0.059 μm for 40 μm spheres and 0.077 μm for 50 μm spheres, corresponding to a relative uncertainty of 0.15% for both the smaller and larger microparticles. These values are 10 times smaller than the uncertainty that we could simply assign from the particle size distributions, propagating to a factor of 30 improvement in volumetric uncertainty upon use of this edge threshold to analyze microdroplet images.

**Edge threshold from dielectric spheres**

We follow the procedures above to calculate dielectric sphere diameter that we use to obtain the edge threshold. To test reproducibility and the dependence of edge threshold on sphere diameter, we measure the diameter and determine the edge threshold of five dielectric spheres, two with nominal diameters of 40 μm and three with nominal diameters of 50 μm, encompassing the range of diameters of our experimental microdroplets. For each dielectric sphere, we run $10^3$ Monte-Carlo trials on each image to obtain the edge threshold distribution (Table S3). We find no variation of the edge threshold between 40 and 50 μm, other than scatter from random effects, indicating that diameters in this range result in similar volume effects in an imaging system with a depth of field ranging from 3 to 5 μm (Figure S5a and S6a). Accordingly, we pool the data from the four dielectric spheres that each have a set of 50 images at three orientations (Table S3). After calculating all four edge threshold distributions, we sample 2500 edge threshold values at random from the results of each to compile a distribution of $10^4$ total values. We use the output distribution, which captures the variations in edge threshold between the dielectric spheres, as an input for the calculation of microdroplet volume.

**Gravimetry calibration and correction**

We calibrate our balance using mass standards that are traceable to the SI through NIST. We select five standards surrounding the mass that we expect to measure of approximately 1.5 g. We handle the mass standards with plastic tweezers and apply a gaseous jet from an aerosol duster to the mass standards before weighing them. We measure each standard twice in random order. We construct a calibration curve that accounts for uncertainty of the mass standards, deviation from a linear relationship, and repeatability of the balance.[10] A parametric bootstrap algorithm calculates the uncertainty for each mass measurement.[10] We apply the calibration curve to the source mass to obtain mass with uncertainty (Table S6). A time correction is necessary because the gravimeter rounds the source time to integer seconds. The correction accounts for the loss of 1 s every 48 source time points to 52 source time points, depending on the data set, and is important because of its direct effect on mean volume (Figure S2d, Scheme S2b).

### Cyclopentanol density measurement

We obtain density data for cyclopentanol as a function of temperature (Figure S12, Table S8) from a NIST measurement service.[11] These data are traceable to the SI and are in good agreement with those of a previous study,[12] building confidence in the results. For this service, we provide an aliquot of our experimental cyclopentanol sample to NIST, and match the temperature range with the potential range of ambient temperature in our microdroplet experiments, from 19 to 23 °C.

In summary of this service,[11] NIST measures fluid density using a vibrating-tube density meter. NIST calibrates the meter by measuring the resonant frequency of the vibrating tube containing ambient air and pure water, and calibrates the densities of the air and water by measuring pressure and temperature. NIST tests the measurements of cyclopentanol density and estimates uncertainty by using the vibrating-tube meter to measure the density of pure water and a NIST Standard Reference Material (SRM) of toluene, comparing the measurement results to the reference values. On the basis of these and previous comparisons, NIST performs a Type B evaluation of uncertainty to conservatively estimate a relative uncertainty of $8 \times 10^{-5}$, corresponding to a 95% coverage interval for a normal distribution, for the measurements of cyclopentanol density. This measurement uncertainty is negligible in comparison to our conservative estimate of the effective uncertainty of cyclopentanol density due to the potential range of ambient temperature in our microdroplet experiments.

### Microscopy object alignment

We orient and position the reference objects and the microdroplet trajectory in the image plane, using kinematic mounts to control orientation and optomechanical stages to control position. We image objects through focus using a servo actuator. Before collecting image data, we verify that each reference object is in approximate alignment with the focal surface and the charge-coupled device (CCD) array. For the aperture array and the double knife edge, we adjust the orientation until the reference object moves through focus in a symmetric pattern near the center of the imaging field. This alignment has little effect, however, because the misalignment out of the $x$–$y$ plane must be large to significantly change pixel size or edge threshold by the corresponding cosine error. For example, an angle of 0.035 rad or 2° out of the $x$–$y$ plane, which would be a gross misalignment in a system with a depth of field from 2 to 5 µm, would result in an apparent pitch or dimensional loss of only 0.1%. To align with the CCD array, we rotate the mount holding the aperture array and double knife edge in the $x$–$y$ plane (Figure 1). For dielectric spheres, we verify that each microparticle is effectively spherical (Dielectric sphere characterization), so that orientation is irrelevant. For microdroplets, we adjust the orientation of the inkjet dispenser so that microdroplets in flight remain in focus.

### Image collection optimization

The light-emitting diode (LED) warms up for at least 1 h before use. During image collection, the pulse length of the function generator turns on the LED for long enough to take advantage of the dynamic range of the CCD (Table S2). The CCD exposure time of 3.938 µs is the shortest possible time, which minimizes blur from microdroplet motion. The CCD delay after inkjet trigger of 5.432 µs maximizes image brightness. We match these parameters for the edge threshold references but not for the aperture array, which requires brighter illumination and a longer exposure time of 163 µs. Similarly, we match dark and light background parameters with each object except for the aperture array, which does not require background correction.

### Number concentrations of single microdroplets

We prepare aqueous suspensions of fluorescent nanoplastics. To filter and degas the dispersion medium, we pump deionized water through a membrane filter of mixed cellulose esters with a pore diameter of 0.22 µm, and hold the water under vacuum for 1.5 h. We perform these processes at least 2 h before jetting and store the water overnight at 20 °C if necessary. Immediately before jetting, we dilute suspensions of fluorescent nanoplastics into the water and gently vortex to mix without adding air bubbles, which could destabilize the jetting. We dilute polystyrene particles with nominal diameters of 200 nm from the original concentration by a factor of $10^4$ to a nominal concentration of approximately $4.5 \times 10^8$ particles/mL. We dilute polystyrene particles with nominal diameters of 500 nm from the original concentration by a factor of $10^3$ to a nominal concentration of approximately $3 \times 10^8$ particles/mL. The nanoplastic particles sorb and carry fluorescent molecules with a peak excitation wavelength of 505 nm and peak emission wavelength of 515 nm. The surfaces of the nanoplastic particles have carboxylate functionalization.

We print microdroplets of the aqueous suspensions of fluorescent nanoplastics onto microscope slides of borosilicate glass, which we use as we receive from the manufacturer. The linear stage carrying the borosilicate substrates receives a trigger from the inkjet drive electronics and then travels at 3 mm/s for 1 s. For jetting at 10 Hz, this stage speed results in a mean spacing of the deposition regions of approximately 0.3 mm.

After deposition and evaporation of the microdroplets, we image nanoplastic particles in each deposition region on the borosilicate surface using an epifluorescence microscope. A halogen lamp illuminates the sample through an objective lens with air immersion, a nominal magnification of 50×, and a numerical aperture of 0.95. The filter set includes an excitation filter with a band-pass from 450 to 490 nm, a dichroic mirror with a long-pass cutoff at 495 nm, and an emission filter with a band-pass from 500 to 550 nm. A matching tube lens focuses images on the sensor of a color CCD camera, with an array of 2752 by 2208 pixels, each with an on-chip size of 4.54 by 4.54 µm. We image the smaller particles with an exposure time of 20 ms and the larger particles with exposure times of 3 and 5 ms. The CCD camera records fluorescence micrographs with an image resolution of approximately 271 nm, on the basis of the Abbe criterion.

To count nanoplastic particles in fluorescence micrographs, we use the Analyze Particles tool in image analysis software.[13] We convert the image from red-green-blue (RGB) color to eight-bit grayscale, manually set the intensity threshold to binarize the image, and digitally count the particles. We visually compare the analysis results to the original image and correct the particle count for a few images in which the thresholding or particle analysis leads to an evident error. For most of the deposition regions, we can readily distinguish between single particles in an agglomerate, because the particles are close to or larger than the resolution limit of the fluorescence microscope and have high enough signal-to-noise ratio to easily identify (Figure S10c). However, for a few of the deposition regions, the likelihood of particle agglomeration normal to the imaging substrate or difficulty distinguishing particles on the imaging surface in an agglomeration yields features with an uncertain count of particles, and we exclude the entire deposition region from the final analysis (Figure S10b). To obtain the mean and uncertainty for the number concentration of a single microdroplet, we perform $10^3$ Monte-Carlo trials, each using the number of particles, a constant, and a different value from the microdroplet volume distribution with the edge threshold from the dielectric spheres. This level of uncertainty evaluation is sufficient for the purpose of the present study, which focuses on the measurand of microdroplet volume rather than the measurand of particle count. However, by including additional evaluation of the uncertainty of counting agglomerate features, possibly by using Poisson occupation statistics and minimum signal-to-noise ratio, this method could measure a more complete sample of a population of single microdroplets and higher values of number concentrations. For 95% of the microdroplets in such a population, particle number will vary by a range over a factor of 2. First droplets are problematic because of incongruity and other nonideality. If a comparison of the particle number concentrations of single microdroplet suspensions to the particle number concentration of a bulk suspension is of interest, then it is possible to reduce random effects in the sampling of single microdroplets by imaging a large number of deposition regions in an array and averaging the results.

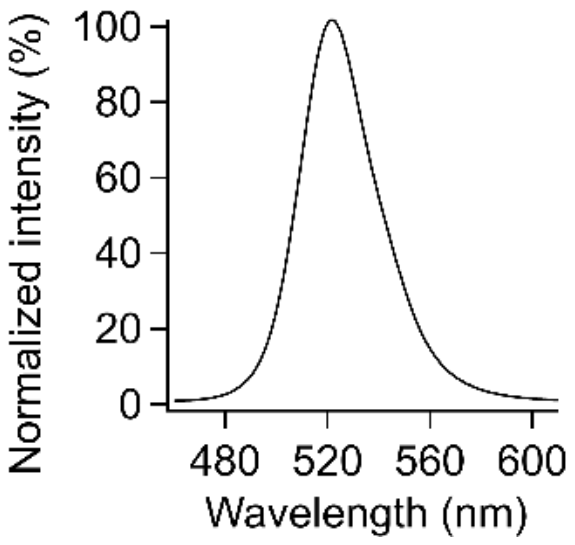

**Figure S1.** Emission spectrum. Plot of the emission spectrum of the illumination LED.

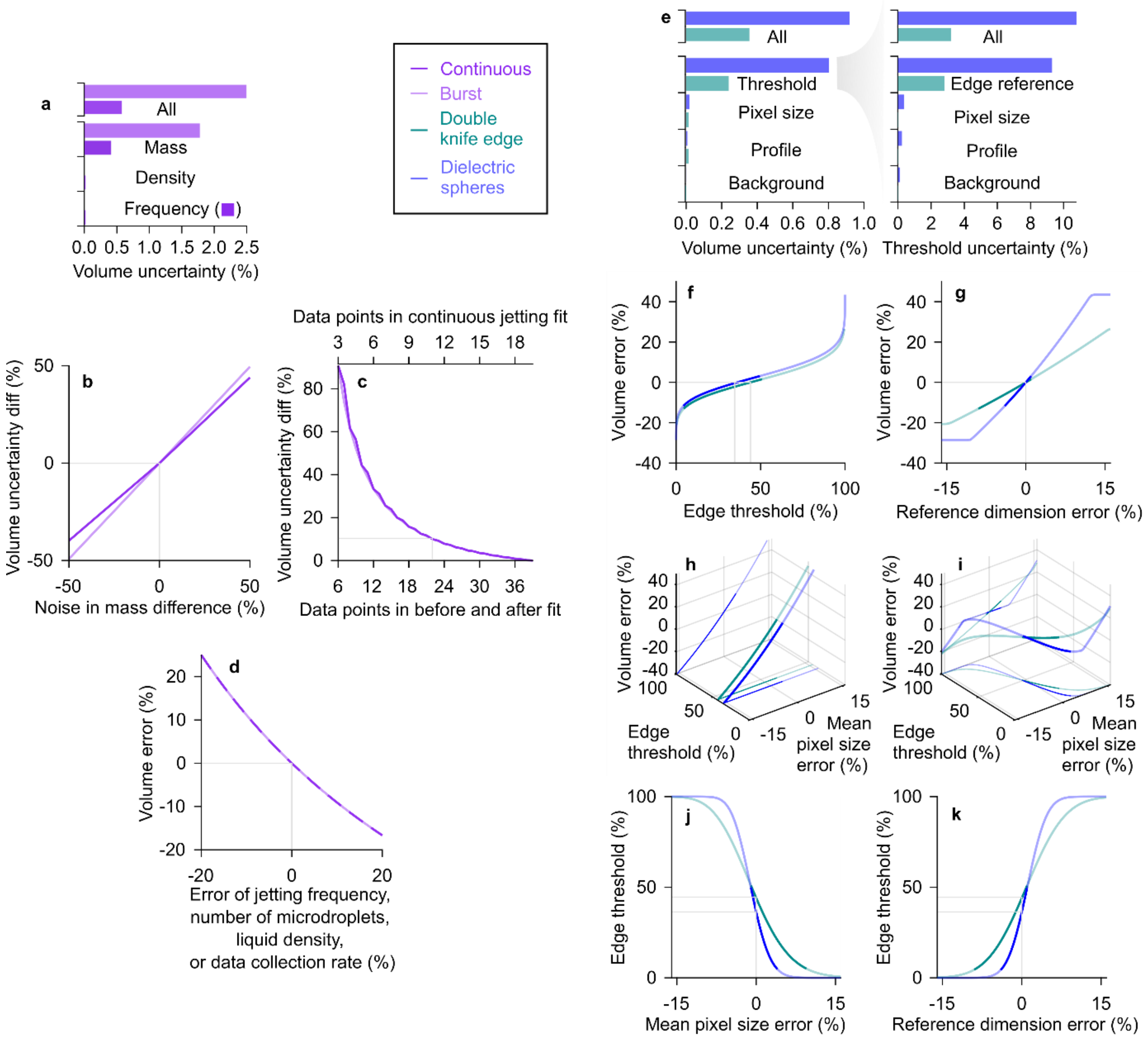

**Figure S2.** Uncertainty and sensitivity analysis. Plots include results from gravimetry with (dark purple) continuous and (light purple) burst mode and microscopy with (teal) double knife edge and (blue) dielectric spheres edge thresholds. Gray lines indicate the values that we use for analysis of microdroplet metrology data in Figure 3. Light portions of lines indicate unreasonable values for microscopy, in regions where the edge threshold is greater than 50% or less than approximately 5%. (a-d) Gravimetry. (a) Bar plot of relative volume uncertainty from all inputs and individual inputs from mass standards and density of jetting liquid. (b) Plot of sensitivity of the volume uncertainty to noise in the mass data. (c) Plot of sensitivity of the volume uncertainty to the number of data points in the fits. (d) Plot of sensitivity of the mean volume to error of jetting frequency, number of microdroplets, liquid density, or data collection rate. (e-k) Microscopy. (e) Bar plots of (left) volume uncertainty from all inputs and individual inputs from edge threshold, pixel size, profile data, and background subtraction, and (right) threshold uncertainty from all inputs and individual inputs from edge reference dimension, pixel size, profile data, and background subtraction. (f) Plot of sensitivity of mean volume to edge threshold. (g) Plot of sensitivity of mean volume to reference dimension error. (h) Plot of sensitivity of mean volume to error in mean pixel size with a constant edge threshold. (i) Plot of sensitivity of mean volume to mean pixel size error with edge threshold variation from the mean pixel size error in (j). (j) Plot of sensitivity of the edge threshold to mean pixel size error. (k) Plot of sensitivity of edge threshold to reference dimension error.

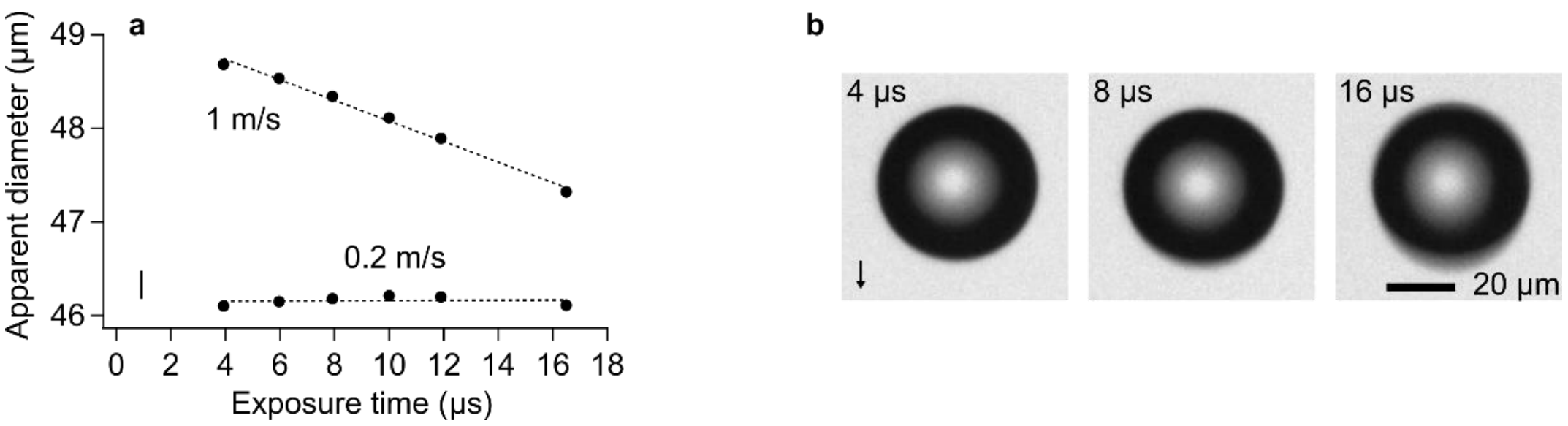

**Figure S3.** Effects of motion blur on apparent diameter. (a) Plot of apparent diameter of microdroplets with constant speeds of 0.2 and 1 m/s and varying exposure time. Plot includes (circles) data, (dash lines) linear fits of data, and (lower left) representative 95% coverage interval of apparent diameter. (b) Transmission micrographs of water microdroplets at exposure times of 4, 8, and 16 µs. The arrow points in the direction of microdroplet motion.

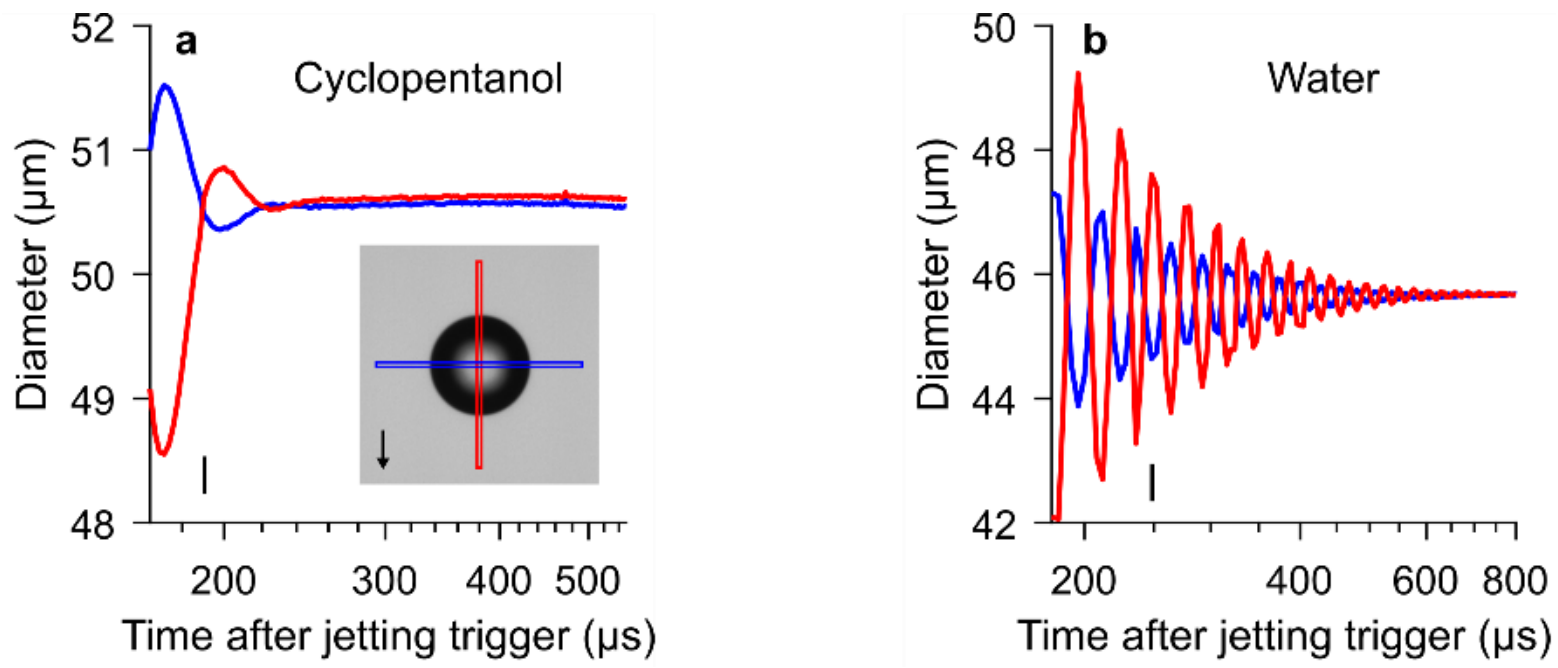

**Figure S4.** Microdroplet oscillation. (a, b) Plots of apparent diameters of microdroplets in flight at the (blue) equator and (red) meridian, immediately after jetting for (a) cyclopentanol with a speed of 0.2 m/s and (b) water with a speed of 0.3 m/s. Both plots include (lower left) representative 95% coverage intervals of apparent diameter. The arrow in the inset transmission micrograph of (a) points in the direction of microdroplet motion.

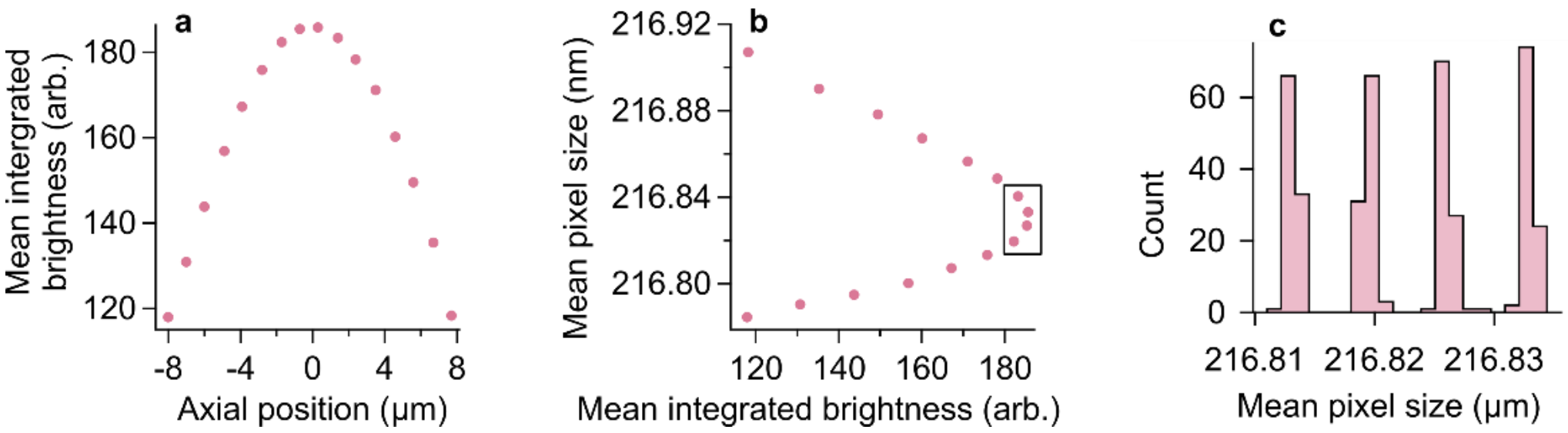

**Figure S5.** Aperture array through focus. Plots (a) and (b) include measurements through focus in axial increments of 1 µm with 100 images at each *z* position. (a) Plot of mean integrated brightness as a function of axial position, with the highest integrated brightness indicating best focus. (b) Plot of mean pixel size as a function of integrated brightness. Each data point in (a) and (b) is the mean value from 100 images. (c) Histogram of mean pixel size from four axial positions near best focus (box in (b)), which we use to construct the uniform distribution for Microdroplet metrology and Number concentrations of single microdroplets.

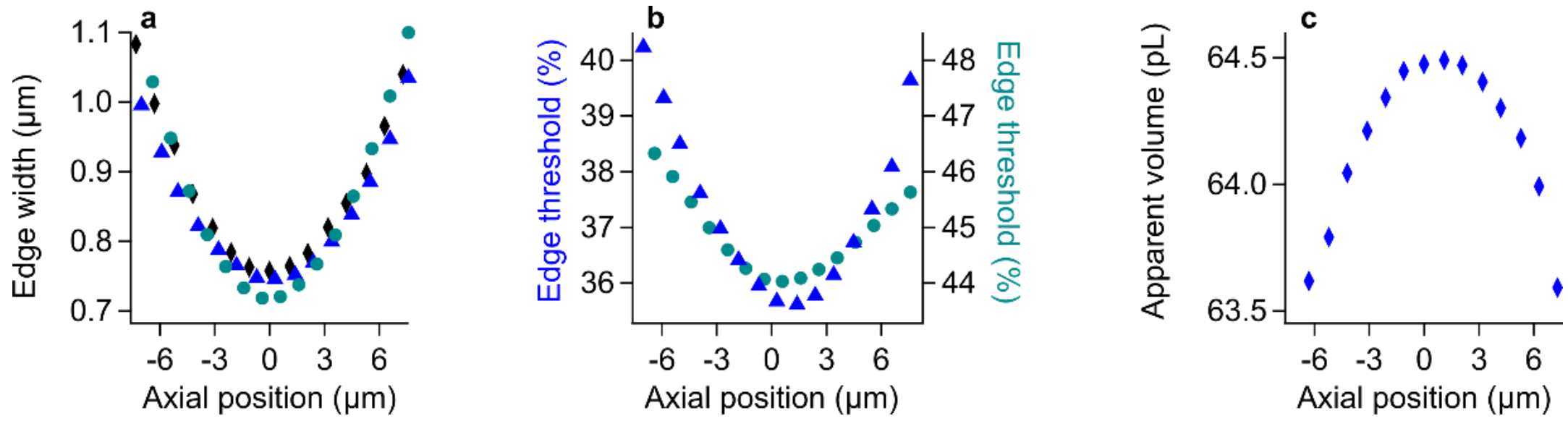

**Figure S6.** Edge references and microdroplets through focus. We manually align the series with edge width minima at axial position of 0 µm. Plots include data in axial increments of 1 µm with 100 images at each *z* position from (teal circle) double knife edge, (blue triangle) dielectric spheres, and (black diamond, blue diamond) microdroplets. (a) Plot of edge widths from imaging a double knife edge, a dielectric sphere, and a microdroplet through focus. (b) Plot of edge thresholds from imaging of a double knife edge and a dielectric sphere through focus. (c) Plot of apparent volume from imaging microdroplets in flight with a speed of 0.4 m/s through focus, using a constant edge threshold from dielectric spheres near best focus.

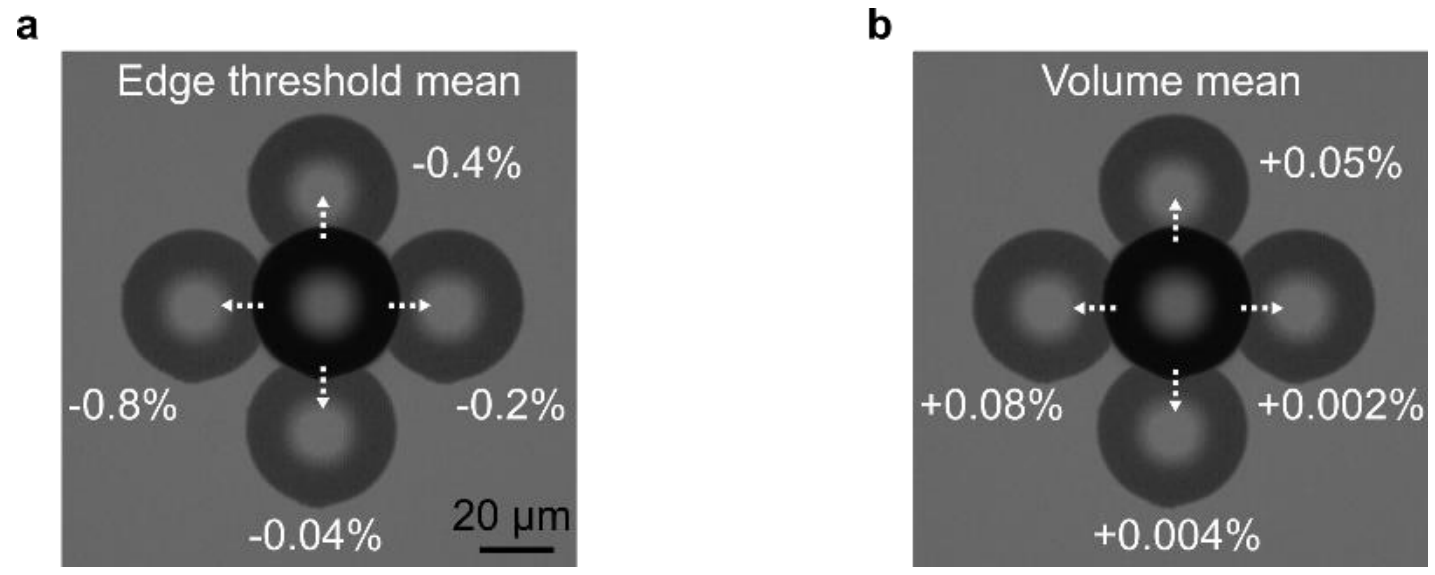

**Figure S7.** Position dependence of edge threshold and volume. (a, b) Overlays of transmission micrographs of the same polystyrene microparticle at five different positions shows the resulting variation of mean values of (a) edge threshold and (b) volume. Normalization is with respect to the center position.

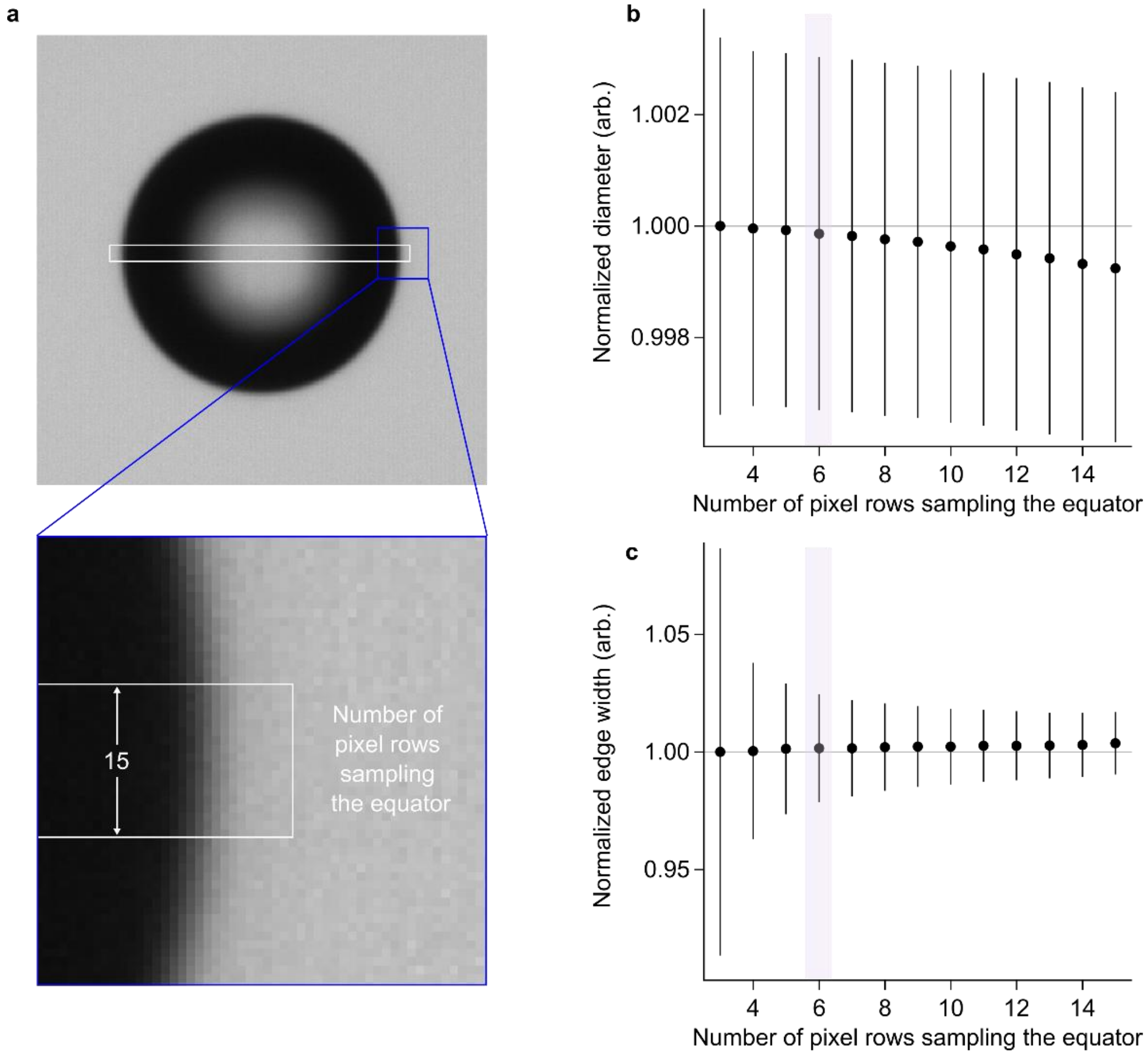

**Figure S8.** Sampling the equator. (a) Transmission micrograph of a cyclopentanol microdroplet with schematic showing the maximum number of pixel rows that we test to sample the equator. The mean pixel size is 216.82 nm with a standard deviation of 0.02 nm. (b) Plot of normalized diameter as a function of the number of pixel rows sampling the equator. (c) Plot of normalized edge width as a function of the number of pixel rows sampling the equator. The data result from 1000 Monte-Carlo trials of 62 cyclopentanol microdroplets with diameters of approximately 50 µm. As the number of pixel rows sampling the equator increases from 3 to 15, (black circles) mean values show systematic trends of decreasing diameter and increasing edge width, with (vertical bars) decreasing relative 95% coverage intervals. The purple shading shows the number of pixel rows that we select to analyze polystyrene microparticles, cyclopentanol microdroplets, and water microdroplets suspending fluorescent nanoplastics. Systematic effects of this selection are insignificant in comparison to the total uncertainty of the measurement.

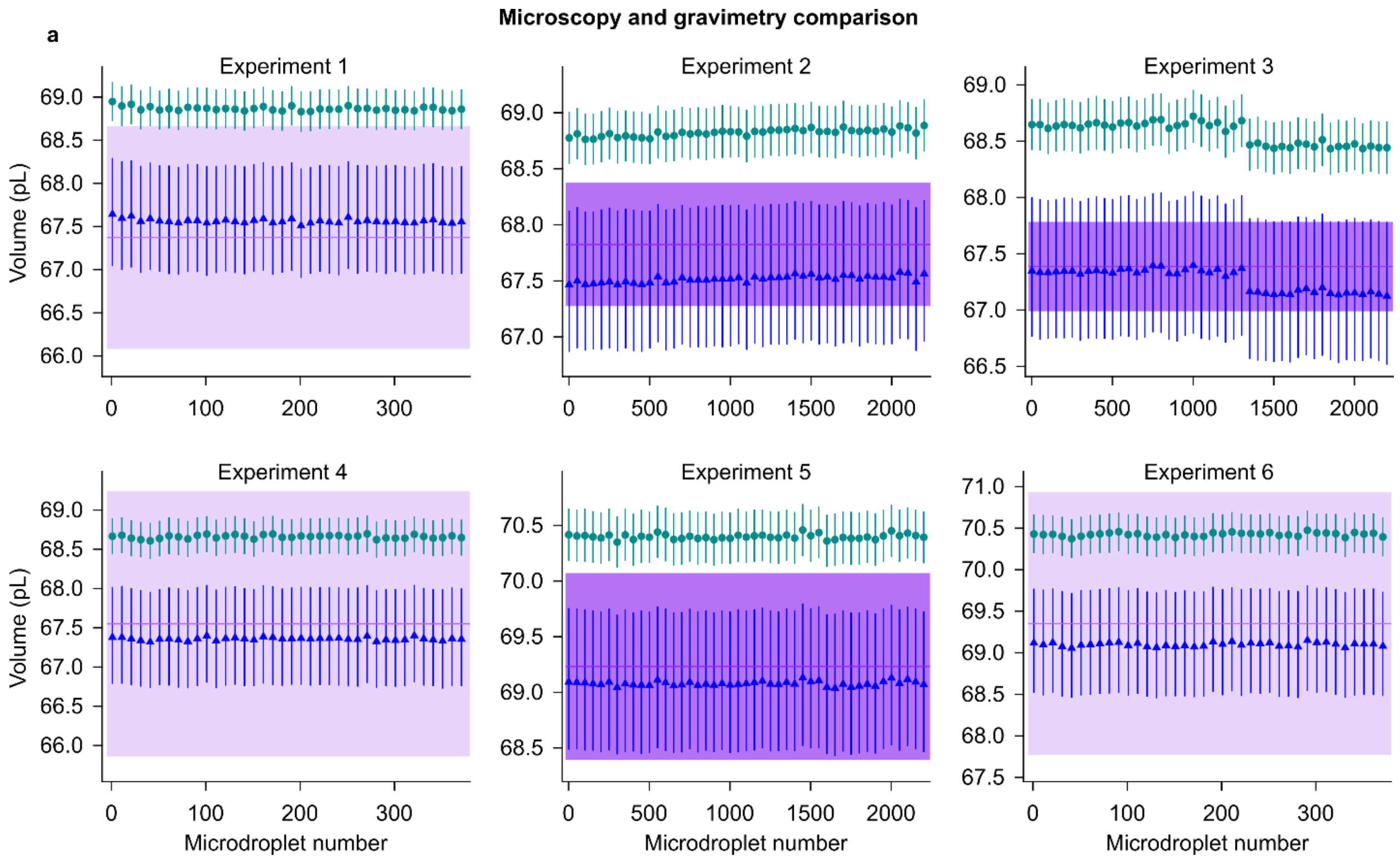

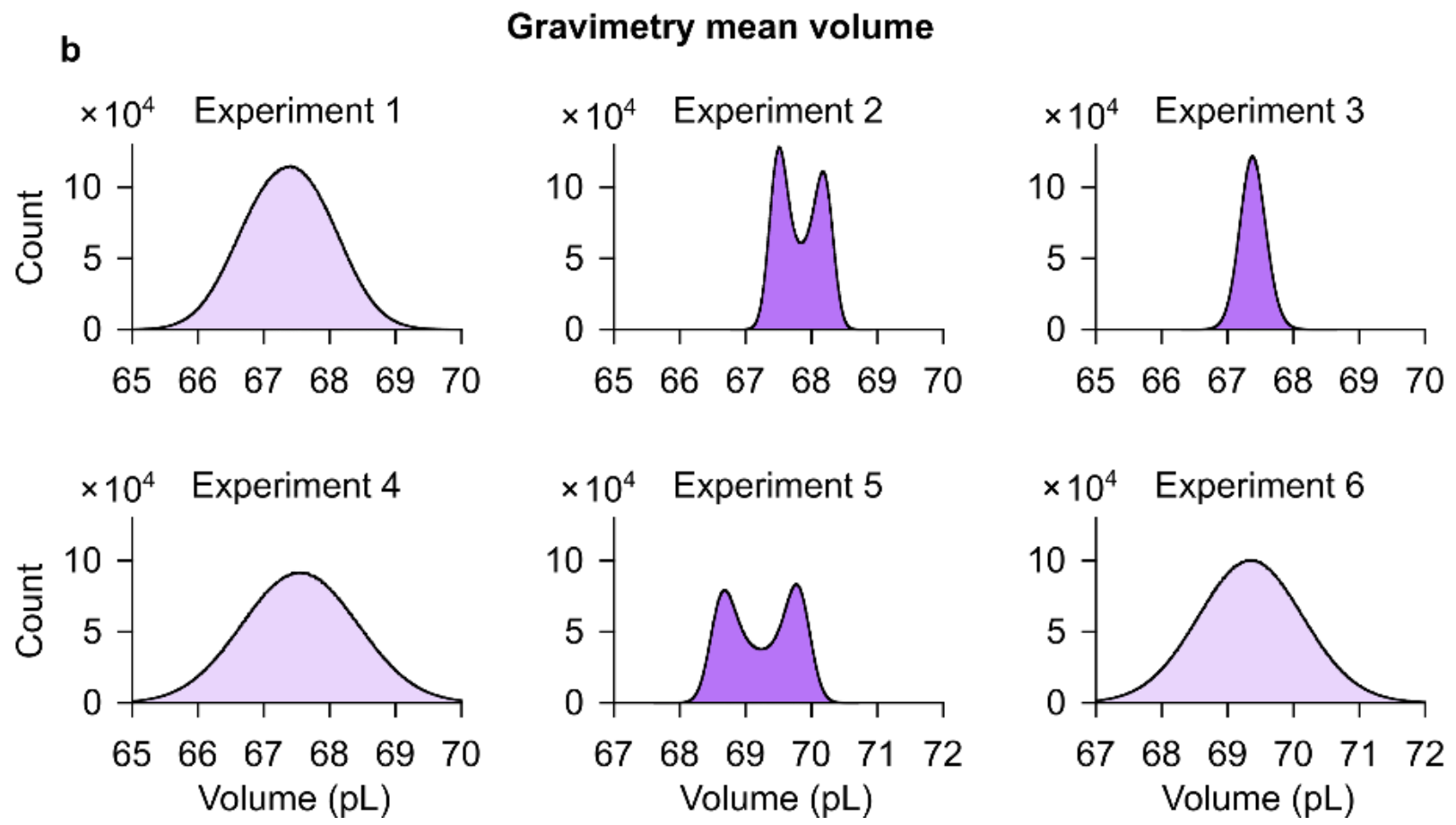

**Figure S9.** Microdroplet metrology results from all measurements. (a) Plots of (purple line and shading) mean volume of a microdroplet ensemble from gravimetry using (light purple) burst and (dark purple) continuous modes, and volume of single microdroplets from microscopy using edge thresholds from analysis of (teal circle) the double knife edge and (blue triangle) dielectric spheres. The microscopy data are for 38 microdroplets for burst mode or 45 microdroplets for continuous mode. Vertical bars are 95% coverage intervals. (b) Histograms of mean volume of microdroplet ensemble from gravimetry with (light purple) burst and (dark purple) continuous mode. Bimodal distributions in Experiments 2 and 5 show the influence of different evaporation rates before and after jetting. Scales of the horizontal axes of the histograms match.

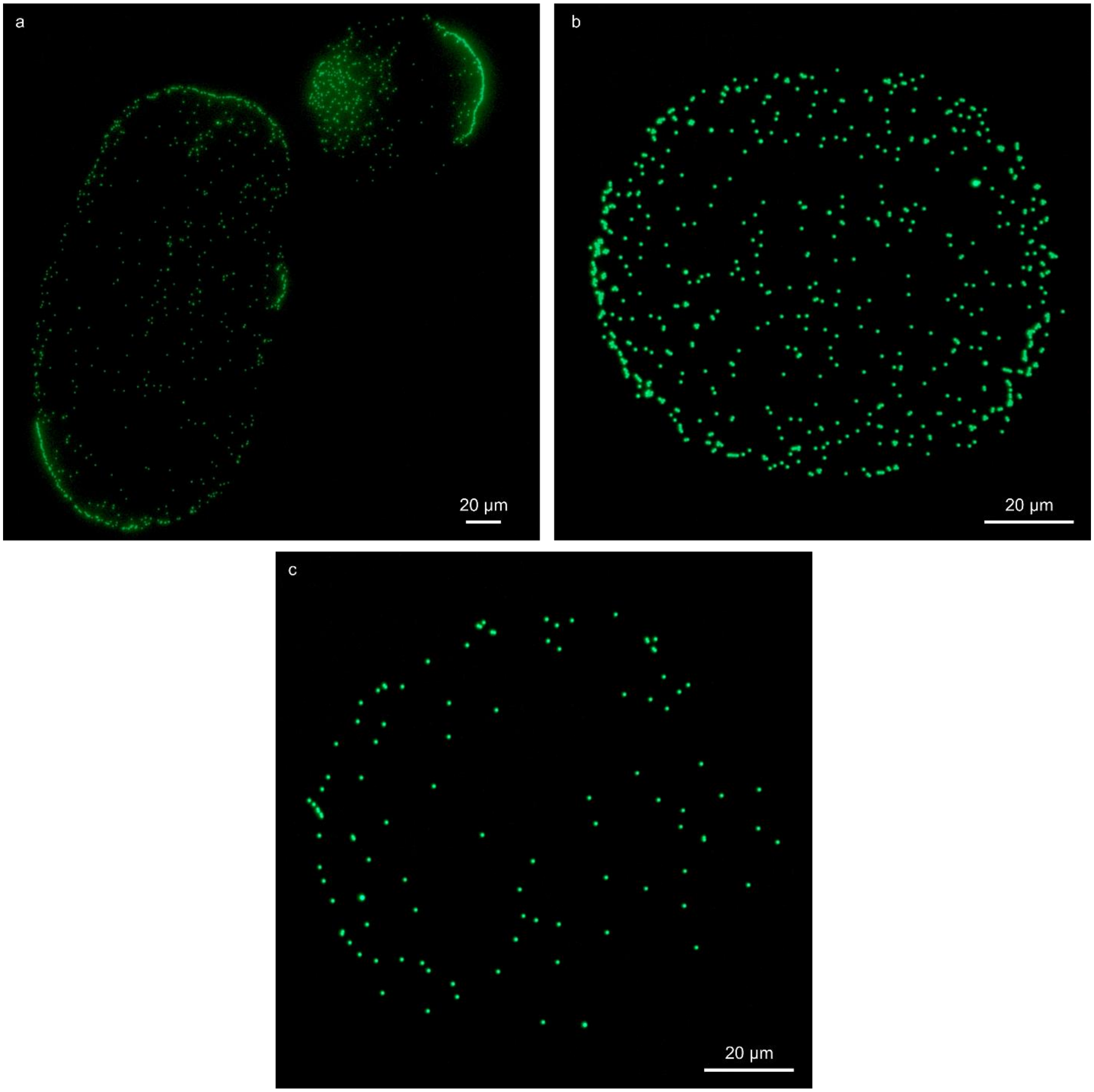

**Figure S10.** Fluorescence micrographs of microdroplet deposition. Brightness and contrast are representative of source data in true color. In (a), the objective lens has a nominal magnification of 20× and a numerical aperture of 0.8. In (b) and (c), the objective lens has a nominal magnification of 50× and a numerical aperture of 0.95. These representative micrographs show (a) deposition of multiple microdroplets in proximity, due to an initial delay of stage translation, which we exclude from analysis, (b) microdroplet deposition with significant agglomeration in which single particles are difficult to distinguish, which we also exclude from analysis, and (c) microdroplet deposition with agglomeration in which single particles are evident, which we include in the analysis.

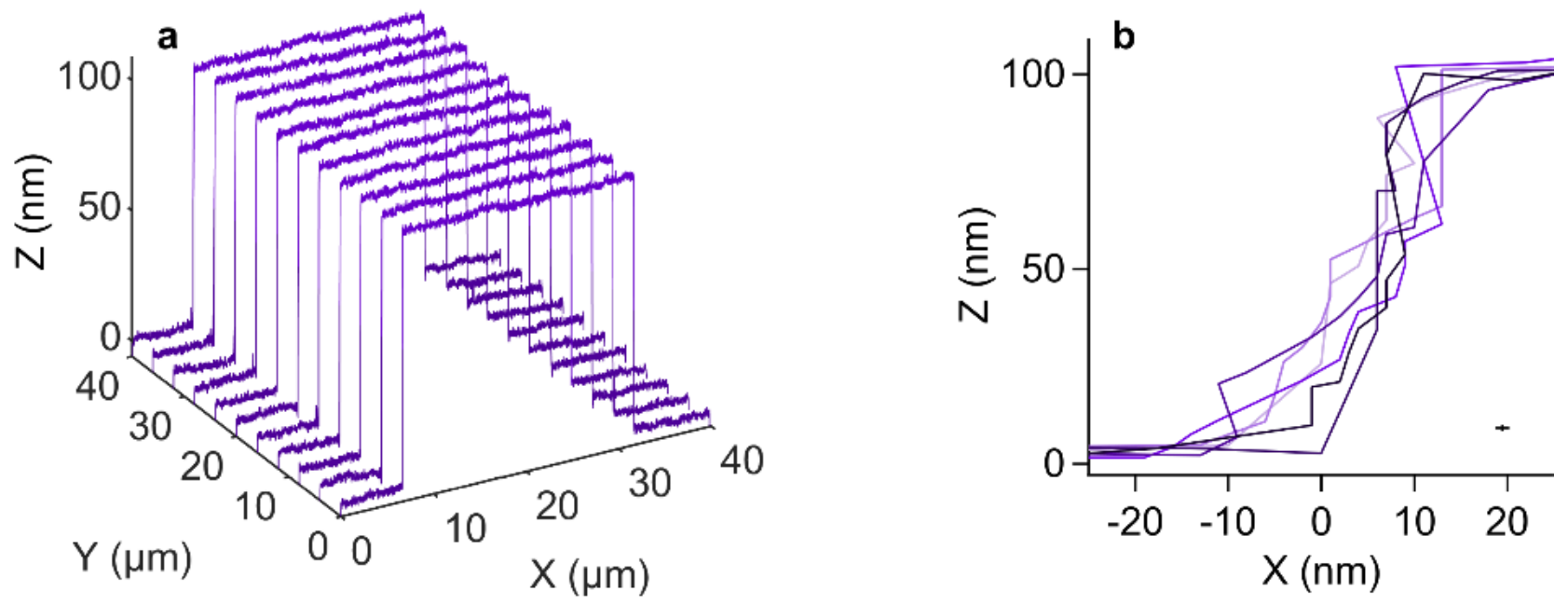

**Figure S11.** Double knife edge characterization. Results from critical-dimension atomic-force microscopy, with a new coordinate system of *X*, *Y*, and *Z*. (a) Three-dimensional plot of surface profiles from a scan area of 40 by 40 µm. (b) Two-dimensional plot magnifying the left knife edge from the same profiles, with the edge position set to approximately zero to clearly show edge width and variation in edge profiles. Plot (b) includes (lower right) representative 95% coverage intervals of the *X* and *Z* measurements.

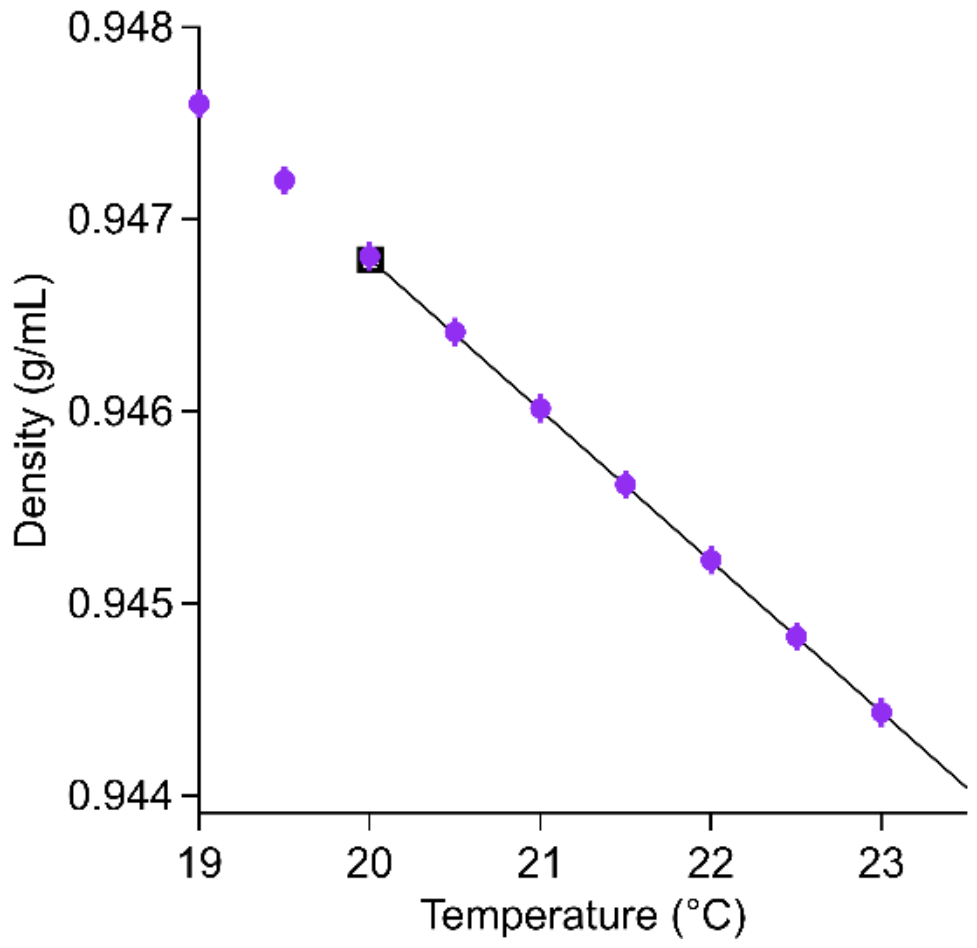

**Figure S12.** Cyclopentanol density as a function of temperature. Plot of (purple circles) NIST measurements[11] of cyclopentanol density and (black square marker and line) data from a previous study.[12] Vertical bars are 95% coverage intervals.

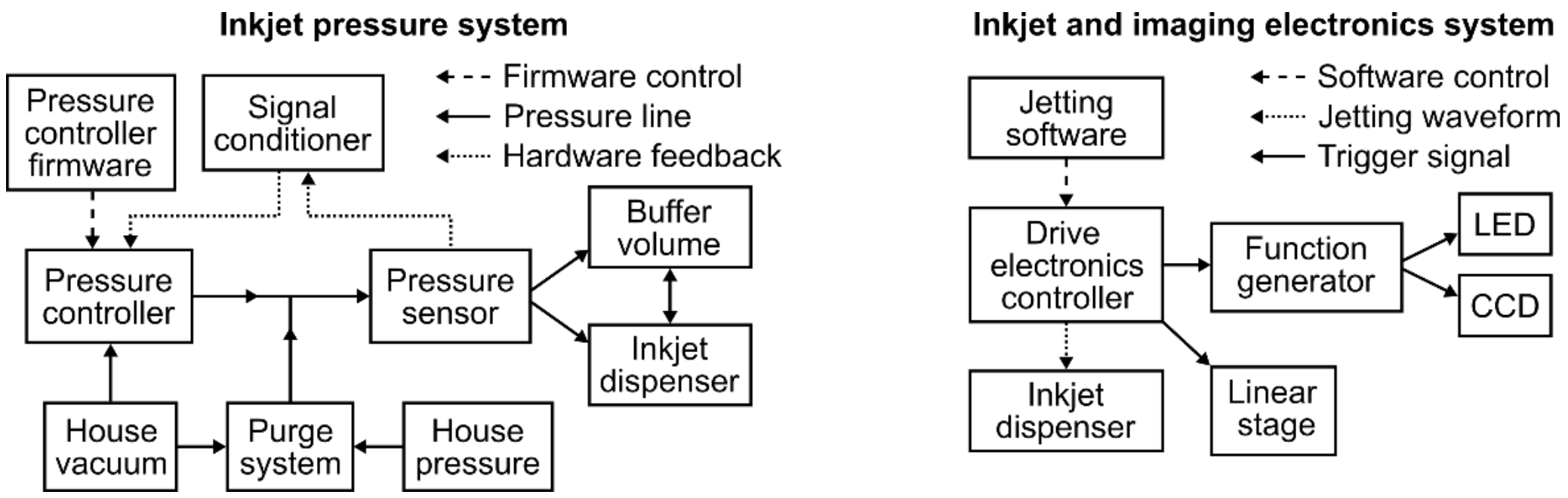

**Scheme S1.** Pressure and electronic systems. These systems supply the pressure, jetting waveform, and trigger signal in Figure 1. Materials are in Table S1. In the inkjet pressure system, to maintain backing pressure, the pressure controller holds the vacuum level set in the pressure controller firmware, with feedback from the pressure sensor through the signal conditioner. The buffer volume of a scintillation vial maintains headspace. The purge system manually loads or empties the inkjet dispenser. In the inkjet and imaging electronics system, the drive electronics controller simultaneously triggers the LED and CCD by a function generator and triggers the linear stage directly. The function generator reshapes the drive electronics output voltage to the pulse length and amplitude that are necessary for the LED and CCD (Table S2).

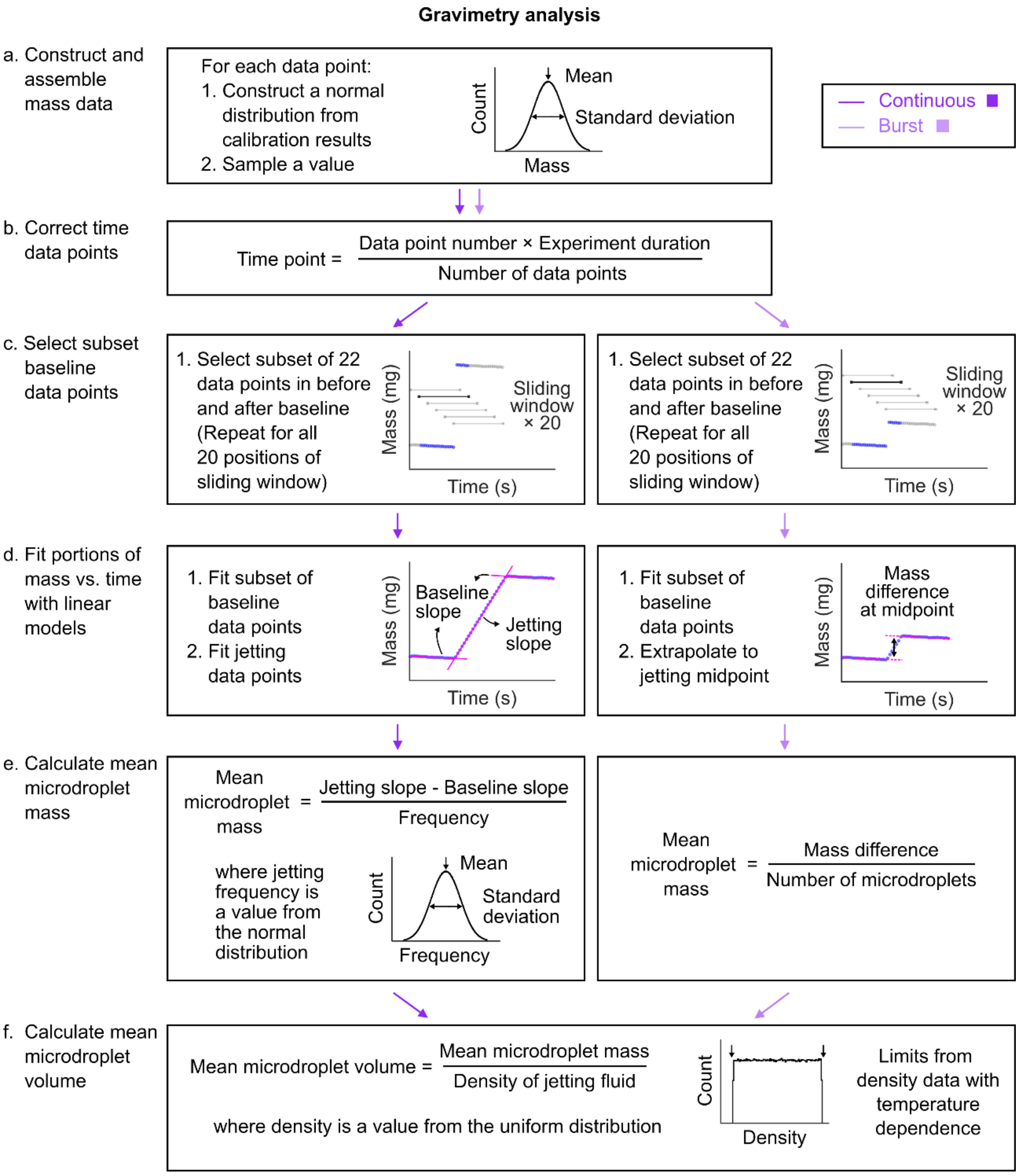

**Scheme S2.** Gravimetry analysis for continuous and burst mode.

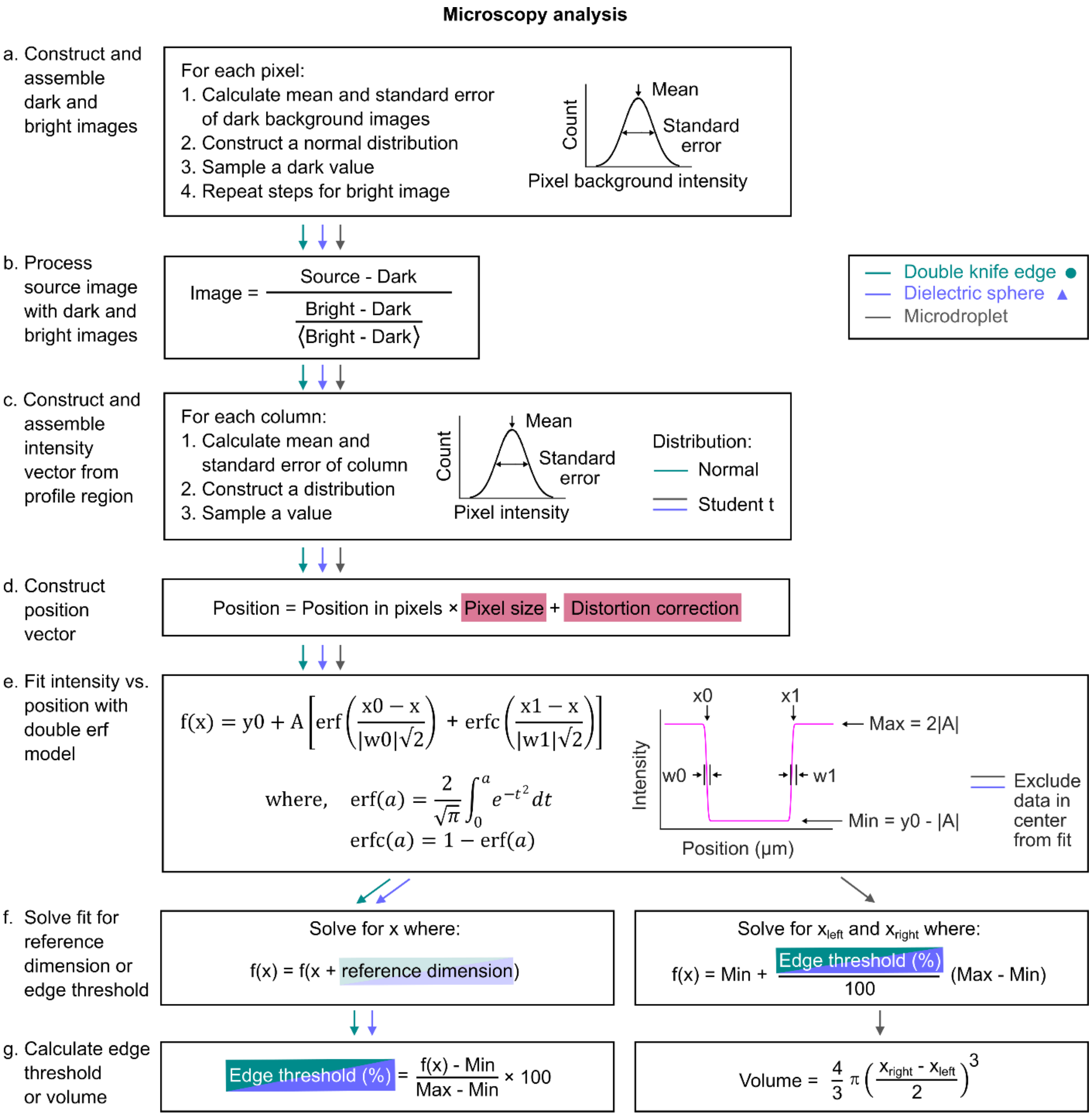

**Scheme S3.** Edge reference and microdroplet image analysis.

**Table S1.** Experimental details of equipment and standards.

| | Item | Details |
|---|---|---|
| Inkjet printing | Inkjet dispenser | Piezoelectric jetting mechanism, cartridge style, orifice inner diameter of 50 μm |
| | Cartridge holder | Inner diameter of slightly greater than 2.5 mL, matches syringe |
| | Syringe | Outer diameter of 2.5 mL, polypropylene, Luer lock, matches cartridge holder |
| | Jetting software | Manual and programmable jetting control, waveform type of 8 point bipolar trapezoidal or sine, pulse mode of single, burst, or continuous, pulse generation control with external transistor–transistor logic (TTL) output, trigger delay range from -500 to 2500 μs relative to jetting |
| | Drive electronics controller | Outputs waveform, universal serial bus (USB) communication with jetting software, external trigger output from 2.5 to 5 V with rising flank of greater than 0.5 μs, jetting frequency range from 1 Hz to 30 kHz, voltage range from -140 to 140 V, rise and fall time range from 1 to 3276 μs, total pulse length of less than 4095 μs, waveform parameter resolution of 1 V and 1 μs |
| | Pressure controller, integrated | Absolute, closed-loop electronic pressure-control system with full-scale range of 133 kPa, maximum pressure differential of 310 kPa, pressure control accuracy of ± 0.2% of full scale, typical response time of less than 1 s, and proportional-integral-derivative (PID) controls |
| | Pressure sensor, differential | Differential pressure transducer with full-scale range of 133 kPa, minimum pressure accuracy of 0.12% of reading, time constant of 40 ms, reference side volume of 25 mL, matches pressure controller |
| | Signal conditioner, high accuracy | Linearizes pressure signal of pressure sensor with analog output from 0 to ±10 V, output linearity of ± (0.005% of reading + 0.001% of full scale), accuracy of 0.01% of reading, display update rate of 1.4 Hz, matches pressure sensor |
| | Purge system | Manufacturer recommendations of input pressure maximum of 410 kPa and input vacuum of 65 kPa, adjustable output pressure |
| | Buffer volume | Scintillation vial with volume of 30 mL and rubber septum |
| Gravimetry | Sub-microgram balance | Readability of 0.1 μg, weighing capacity of 2.1 g, repeatability of ≤ ± 0.25 μg, linearity of ≤ ± 0.9 μg, response time of ≤ 10 s, display update rate from 0.2 to 0.4 s |
| | Gravimetric analysis software | Custom functions by authors and others[14] |
| | Mass standards | Table S6 |
| Microscopy equipment | LED | Power output of 1 W after collimation, Emission spectrum (Figure S1) |
| | Condensing lens | Plano-convex, diameter of 25.4 mm, focal length of 25.4 mm, antireflective coating for the wavelength range from 350 to 700 nm, refractive index of 1.515 |
| | Objective lens | Nominal magnification of 20×, numerical aperture of 0.22, working distance of 12.1 mm, air immersion, infinity correction, matches tube lens |
| | Tube lens | Focal length of 165 mm, working distance of 118 mm, antireflective coating for the wavelength range from 350 to 700 nm, matches objective lens |
| | CCD camera | CCD sensor array of 1600 by 1200 pixels, on-chip pixel size of 4.4 by 4.4 μm, analog-to-digital output of 8 bits, maximum imaging rate of 15 Hz with full field, minimum exposure time of 3.938 μs, TTL input and output triggers |
| | CCD software | Constant shutter control, single image and video capture modes, manual or trigger modes of image collection, controllable exposure delay |
| | Function generator | Pulse waveform output with frequency range from 1 mHz to 25 MHz, TTL output, rise/fall time of ≤ 9 ns |
| | Actuator | Direct current servo motor, absolute cyclic pitch error of ± 0.5 μm |
| | Actuator controller | Minimum increment of *z* position of approximately 1 μm |
| | Image analysis software | Custom functions by authors and others[15–17] |
| Microscopy standards | Aperture arrays | Platinum film with approximate thickness of 80 nm, on titanium adhesion layer with thickness of approximately 5 nm, on silica substrate with thickness of 0.17 mm, Table S3 |
| | Double knife edge | Chromium film with an optical density of 3 at a wavelength of 450 nm with thickness approximately 100 nm, on quartz substrate with thickness of 2.3 mm (5009 photomask), Table S3 |
| | Dielectric spheres | Polystyrene, refractive index of 1.59 at a wavelength of 589 nm. Nominal diameter of 40 μm has a mean diameter of 39.94 ± 0.35 μm with standard deviation of 0.35 μm. Nominal diameter of 50 μm has a mean diameter of 50.2 ± 0.3 μm with standard deviation of 0.5 μm. Uncertainty of the mean value is the measurement uncertainty plus the standard deviation of at least nine measurements from the product datasheet, Table S3 |
| | CD-AFM | References[8,9] |
| | Optical profilometer | Measurement array of 1232 by 1028 pixels, white-light LED illumination source with peak wavelength of 550 nm, vertical scanning interferometer (VSI) mode, maximum vertical scanning range of 10 mm, vertical resolution of 1 nm, imaging with nominal magnification of 10×, numerical aperture of 0.3, and spatial sampling of 1.38 µm, or nominal magnification of 50×, numerical aperture of 0.55, and spatial sampling of 0.28 µm, *z* stage resolution of 2 nm with linear stage and 0.75 nm with piezo stage |
| | Surface metrology software | Visualization of three-dimensional surface topography, leveling, stitching of multiple overlapping measurements from a horizontal plane into a single surface, two-dimensional profile extraction and measurement of step height |
| | Step height standards | Chromium films on quartz and gauge blocks on optical flats, Table S4 |

**Table S2.** Function generator settings for imaging during jetting.

| Dataset | Mode | Inkjet trigger frequency (Hz) | Pulse shape | Pulse length (µs) | Peak to peak amplitude (V) | Offset (V) | Duty cycle (%) |
|---|---|---|---|---|---|---|---|
| Microdroplet metrology | Burst pulse | 10 | Square wave | 100 | 3.6 | 2.5 | 50 |
| Microdroplet deposition | Burst pulse | 10 | Square wave | 125 | 3.6 | 2.5 | 50 |

**Table S3.** Characterization results for microscopy standards.[a-c]

| Standard | Reference value | Figure or scheme | Distribution | Distribution mean or limits | Standard deviation | Data source |
|---|---|---|---|---|---|---|
| Aperture array | Array pitch | Figure 2f | Normal[b] | Mean: 5.001 45 µm | 0.000 54 µm | Reference[4] |
| Double knife edge | Edge separation | Figure 2j Scheme S3f | Uniform[b] | Lower: 24.955 µm<br>Upper: 25.046 µm | $3.410 \times 10^{-3}$ µm<br>$4.485 \times 10^{-3}$ µm | CD-AFM |
| Dielectric sphere 1 | Sphere diameter | Figure 2n Scheme S3f | Normal[c] | Mean: 50.00$\underline{1}$ µm | 0.07$\underline{7}$ µm | Optical profilometry |
| Dielectric sphere 2 | | | | Mean: 39.24$\underline{0}$ µm | 0.05$\underline{9}$ µm | |
| Dielectric sphere 3 | | | | Mean: 50.20$\underline{6}$ µm | 0.07$\underline{7}$ µm | |
| Dielectric sphere 4 | | | | Mean: 50.16$\underline{3}$ µm | 0.07$\underline{7}$ µm | |

An underline indicates a non-significant figure to avoid a rounding error.
[a]Abbreviation: CD-AFM = critical-dimension atomic-force microscopy.
[b]We make a subject-matter judgement to estimate the distribution type.
[c]Maximum-likelihood estimates from calibration asymptotically follow a normal distribution.

**Table S4.** Step height standards for optical profilometry calibration.

| Reference step height, mean (µm) | Reference step height, standard deviation (µm) | Source step height, mean (µm) |
|---|---|---|
| 0.9542 | 0.0024 | 0.9827, 0.9780 |
| 10.139 | 0.034 | 10.15, 10.23 |
| 19.426 | 0.051 | 19.55, 19.55 |
| 22.90 | 0.06 | 22.91, 22.78 |
| 152.37 | 0.34 | 152.6, 152.3 |

The mean values and standard deviations of reference step height are from the corresponding calibrations.
The mean values of source step height are from our optical profilometry measurements.

**Table S5.** Electronic and pressure settings for inkjet printing.

| Dataset | Jetting fluid | Micro-droplet speed (m/s) | Backing pressure (Pa) | Wave-form shape | Wave-form period (µs) | Wave-form amplitude (V) | Jetting frequency (Hz) | Inkjet trigger divider | Inkjet trigger delay (µs) |
|---|---|---|---|---|---|---|---|---|---|
| Lower velocity metrology | Cyclo-pentanol | 0.2 | 800 | Sine | 92 | 46 | 100 | 10 | 500 |
| Higher velocity metrology | Cyclo-pentanol | 0.4 | 800 | Sine | 91 | 46 | 100 | 10 | 500 |
| Micro-droplet deposition | Water with nanoplastics | 0.3 | 1750 | Sine | 89 | 25 | 10 | 1 | 800 |

**Table S6.** Mass standards for gravimetry calibration.

| Reference mass (g) | 95% confidence interval (g) | Source mass (g) |
|---|---|---|
| 2.000 002 1 | ± 0.000 006 0 | 2.000 002, 2.000 002 |
| 0.999 999 9 | ± 0.000 005 0 | 1.000 006, 1.000 000 |
| 0.500 001 4 | ± 0.000 002 5 | 0.499 994, 0.499 997 |
| 0.200 000 3 | ± 0.000 002 1 | 0.199 994, 0.199 991 |
| 0.100 000 5 | ± 0.000 002 1 | 0.999 990, 0.100 001 |

Reference masses and 95% confidence intervals are from the calibration certificate.
We obtain source mass values by weighing each reference mass twice in random order.

**Table S7.** Distributions for Monte-Carlo evaluation of uncertainty.[a-d]

| | Parameter | Figure or scheme | Distribution | Distribution mean or limits and uncertainty | Parameter or data source |
|---|---|---|---|---|---|
| Gravimetry | Gravimetric mass | Figure 2a<br>Scheme S2a | Normal[c] | For each data point:<br>Mean = mass after calibration<br>Standard deviation = standard deviation from calibration | Table S10 |
| | Cyclopentanol density | Figure 2b<br>Scheme S2f | Uniform[b] | Lower limit = density at 22.5 °C = 0.9448 g/mL<br>Upper limit = density 19.5 °C = 0.9472 g/mL | Table S8 |
| | Frequency (continuous mode) | Scheme S2e | Normal[b] | Mean = 100.009 Hz<br>Standard deviation = 0.0051 Hz | Oscilloscope measurement and specification |
| | Microdroplet volume | Figure 2c<br>Scheme S2f | Analysis output | Varies by jetting mode and experiment | Analysis of mass data |
| Microscopy | Mean pixel size in the *x* direction | Figure 2g | Normal[b] | Mean = 216.82 nm<br>Standard deviation = 0.02 nm | Analysis of aperture array through focus |
| | Pixel size distortion correction in the *x* direction | Figure 2g | Analysis output, map | Field dependent. Single-pixel correction distances range from -25 to 65 nm with a mean of 23 nm | |
| | Distortion correction uncertainty | Scheme S3d | Normal[b] | For each pixel:<br>Mean = 0 nm<br>Standard deviation = 3.1 nm | |
| | Dark background | Scheme S3a | Normal[d] | For each pixel:<br>Mean = mean across all images (mean across full field is 0.1 ADU)<br>Standard error = standard error across all images (mean across full field is 0.3 ADU) | 1000 light and 1000 dark background images |
| | Bright background | Scheme S3a | Normal[d] | For each pixel:<br>Mean = mean across all images (mean across full field is 190 ADU)<br>Standard error = standard error across all images (mean across full field is 4 ADU) | |
| | Double knife edge intensity profile | Figure 2i<br>Scheme S3c | Normal[d] | For each column of pixels:<br>Mean = mean of intensity across 177 rows of pixels<br>Standard error = standard error across 177 rows of pixels | Image after background subtraction and normalization Figure 2h, l, p Scheme S3b |
| | Dielectric sphere and microdroplet intensity profile | Figure 2m, q<br>Scheme S3c | Student t[d] | For each column of pixels:<br>Mean = mean of 6 rows of pixels at widest point of equator<br>Standard error = standard error of 6 rows of pixels at widest point of equator | |
| | Edge threshold, double knife edge | Figure 2j<br>Scheme S3g | Analysis output | Mean = 44.04%<br>Standard deviation = 0.80% | Edge threshold analysis of in-focus images |
| | Edge threshold, dielectric spheres | Figure 2n<br>Scheme S3g | Analysis output | Mean = 35.94%<br>Standard deviation = 1.91% | |
| | Microdroplet volume | Figure 2s<br>Scheme S3g | Analysis output | Varies by edge threshold and experiment | Volume analysis of in-focus images |

[a]Abbreviation: ADU = analog-to-digital unit.
[b]We make a subject-matter judgement to estimate the distribution type.
[c]Maximum-likelihood estimates from calibration asymptotically follow a normal distribution.
[d]For pixel averages, the central limit theorem applies, indicating an approximately normal distribution for 30 or more replicate measurements or a Student t distribution for less than 30 replicate measurements.

**Table S8.** Cyclopentanol density as a function of temperature.[11]

| Temperature (°C) | Density (g/mL) | Standard deviation (g/mL) |
|---|---|---|
| 19 | 0.947 601 | 0.000 038 |
| 19.5 | 0.947 203 | 0.000 038 |
| 20 | 0.946 809 | 0.000 038 |
| 20.5 | 0.946 413 | 0.000 038 |
| 21 | 0.946 016 | 0.000 038 |
| 21.5 | 0.945 620 | 0.000 038 |
| 22 | 0.945 226 | 0.000 038 |
| 22.5 | 0.944 829 | 0.000 038 |
| 23 | 0.944 433 | 0.000 038 |

An underline indicates a non-significant figure.

**Table S9.** Volume results from microdroplet metrology.

| | Experiment number and mode | Mean volume (pL) | Volume 95% coverage interval | | Difference of mean from gravimetry | | Mean diameter (µm) | Diameter 95% coverage interval | |
|---|---|---|---|---|---|---|---|---|---|
| | | | (pL) | (%) | (pL) | (%) | | (µm) | (%) |
| Gravimetry | 1, Burst | 67.37 | ± 1.29 | ± 1.92 | -- | -- | -- | -- | -- |
| | 2, Continuous | 67.82 | ± 0.55 | ± 0.81 | -- | -- | -- | -- | -- |
| | 3, Continuous | 67.39 | ± 0.40 | ± 0.59 | -- | -- | -- | -- | -- |
| | 4, Burst | 67.55 | ± 1.69 | ± 2.50 | -- | -- | -- | -- | -- |
| | 5, Continuous | 69.23 | ± 0.84 | ± 1.21 | -- | -- | -- | -- | -- |
| | 6, Burst | 69.35 | ± 1.58 | ± 2.28 | -- | -- | -- | -- | -- |
| | Root mean square | Burst | ± 1.52 | ± 2.23 | | | | | |
| | | Cont. | ± 0.60 | ± 0.87 | | | | | |
| Microscopy with dielectric spheres | 1, Burst | 67.56 | ± 0.63 | ± 0.93 | 0.19 | 0.28 | 50.53 | ± 0.15 | ± 0.30 |
| | 2, Continuous | 67.51 | ± 0.63 | ± 0.93 | -0.31 | -0.46 | 50.52 | ± 0.15 | ± 0.31 |
| | 3, Continuous | 67.27 | ± 0.62 | ± 0.93 | -0.12 | -0.17 | 50.46 | ± 0.16 | ± 0.32 |
| | 4, Burst | 67.36 | ± 0.62 | ± 0.93 | -0.19 | -0.29 | 50.48 | ± 0.15 | ± 0.30 |
| | 5, Continuous | 69.07 | ± 0.63 | ± 0.92 | -0.16 | -0.23 | 50.91 | ± 0.15 | ± 0.30 |
| | 6, Burst | 69.10 | ± 0.63 | ± 0.92 | -0.25 | -0.36 | 50.91 | ± 0.15 | ± 0.30 |
| | Root mean square | | ± 0.63 | ± 0.92 | 0.21 | 0.31 | | ± 0.16 | ± 0.31 |
| Microscopy with double knife edge | 1, Burst | 68.86 | ± 0.23 | ± 0.34 | 1.49 | 2.21 | 50.85 | ± 0.06 | ± 0.12 |
| | 2, Continuous | 68.82 | ± 0.23 | ± 0.34 | 0.99 | 1.47 | 50.84 | ± 0.06 | ± 0.13 |
| | 3, Continuous | 68.57 | ± 0.23 | ± 0.34 | 1.19 | 1.76 | 50.78 | ± 0.08 | ± 0.16 |
| | 4, Burst | 68.66 | ± 0.23 | ± 0.34 | 1.11 | 1.64 | 50.80 | ± 0.06 | ± 0.12 |
| | 5, Continuous | 70.39 | ± 0.23 | ± 0.33 | 1.16 | 1.68 | 51.23 | ± 0.06 | ± 0.12 |
| | 6, Burst | 70.42 | ± 0.23 | ± 0.33 | 1.07 | 1.55 | 51.24 | ± 0.06 | ± 0.12 |
| | Root mean square | | ± 0.23 | ± 0.33 | 1.18 | 1.73 | | ± 0.07 | ± 0.13 |

Coverage interval value is the average of the upper and lower 95% coverage interval values.

**Table S10.** Gravimetric data from microdroplet metrology.[e]

| Experiments 1 and 2 | | | | |
|---|---|---|---|---|
| Source time (h:min:s) | Source mass (mg) | Time (s) | Mass (mg) | Standard deviation (mg) |
| 9:09:03 | 1606.795 | 0.0000 | 1606.795 | 0.002 158 40 |
| 9:09:13 | 1606.792 | 9.9808 | 1606.792 | 0.002 158 40 |
| 9:09:23 | 1606.790 | 19.9615 | 1606.790 | 0.002 158 40 |
| 9:09:33 | 1606.787 | 29.9423 | 1606.787 | 0.002 158 40 |
| 9:09:43 | 1606.785 | 39.9231 | 1606.785 | 0.002 158 40 |
| 9:09:53 | 1606.782 | 49.9038 | 1606.782 | 0.002 158 40 |
| 9:10:03 | 1606.780 | 59.8846 | 1606.780 | 0.002 158 40 |
| 9:10:13 | 1606.778 | 69.8654 | 1606.778 | 0.002 158 40 |
| 9:10:23 | 1606.775 | 79.8462 | 1606.775 | 0.002 158 40 |
| 9:10:33 | 1606.773 | 89.8269 | 1606.773 | 0.002 158 40 |
| 9:10:43 | 1606.771 | 99.8077 | 1606.771 | 0.002 158 40 |
| 9:10:53 | 1606.769 | 109.7885 | 1606.769 | 0.002 158 40 |
| 9:11:03 | 1606.767 | 119.7692 | 1606.767 | 0.002 158 40 |
| 9:11:13 | 1606.765 | 129.7500 | 1606.765 | 0.002 158 40 |
| 9:11:23 | 1606.763 | 139.7308 | 1606.763 | 0.002 158 40 |
| 9:11:33 | 1606.761 | 149.7115 | 1606.761 | 0.002 158 40 |
| 9:11:43 | 1606.759 | 159.6923 | 1606.759 | 0.002 158 40 |
| 9:11:53 | 1606.757 | 169.6731 | 1606.757 | 0.002 158 40 |
| 9:12:02 | 1606.755 | 179.6538 | 1606.755 | 0.002 158 40 |
| 9:12:12 | 1606.753 | 189.6346 | 1606.753 | 0.002 158 40 |
| 9:12:22 | 1606.751 | 199.6154 | 1606.751 | 0.002 158 40 |
| 9:12:32 | 1606.749 | 209.5962 | 1606.749 | 0.002 158 40 |
| 9:12:42 | 1606.747 | 219.5769 | 1606.747 | 0.002 158 40 |
| 9:12:52 | 1606.745 | 229.5577 | 1606.745 | 0.002 158 40 |
| 9:13:02 | 1606.743 | 239.5385 | 1606.743 | 0.002 158 40 |
| 9:13:12 | 1606.741 | 249.5192 | 1606.741 | 0.002 158 40 |
| 9:13:22 | 1606.739 | 259.5000 | 1606.739 | 0.002 158 40 |
| 9:13:32 | 1606.737 | 269.4808 | 1606.737 | 0.002 158 40 |
| 9:13:42 | 1606.735 | 279.4615 | 1606.735 | 0.002 158 40 |
| 9:13:52 | 1606.733 | 289.4423 | 1606.733 | 0.002 158 40 |
| 9:14:02 | 1606.731 | 299.4231 | 1606.731 | 0.002 158 40 |
| 9:14:12 | 1606.750 | 309.4038 | 1606.750 | 0.002 158 40 |
| 9:14:22 | 1606.810 | 319.3846 | 1606.810 | 0.002 158 40 |
| 9:14:32 | 1606.870 | 329.3654 | 1606.870 | 0.002 158 40 |
| 9:14:42 | 1606.931 | 339.3462 | 1606.931 | 0.002 158 40 |
| 9:14:52 | 1606.976 | 349.3269 | 1606.976 | 0.002 158 40 |
| 9:15:02 | 1606.975 | 359.3077 | 1606.975 | 0.002 158 40 |
| 9:15:12 | 1606.973 | 369.2885 | 1606.973 | 0.002 158 40 |
| 9:15:22 | 1606.971 | 379.2692 | 1606.971 | 0.002 158 40 |
| 9:15:32 | 1606.969 | 389.2500 | 1606.969 | 0.002 158 40 |
| 9:15:42 | 1606.968 | 399.2308 | 1606.968 | 0.002 158 40 |
| 9:15:52 | 1606.966 | 409.2115 | 1606.966 | 0.002 158 40 |
| 9:16:02 | 1606.964 | 419.1923 | 1606.964 | 0.002 158 40 |
| 9:16:12 | 1606.962 | 429.1731 | 1606.962 | 0.002 158 40 |
| 9:16:22 | 1606.960 | 439.1538 | 1606.960 | 0.002 158 40 |
| 9:16:32 | 1606.959 | 449.1346 | 1606.959 | 0.002 158 40 |
| 9:16:42 | 1606.957 | 459.1154 | 1606.957 | 0.002 158 40 |
| 9:16:52 | 1606.956 | 469.0962 | 1606.956 | 0.002 158 40 |
| 9:17:02 | 1606.954 | 479.0769 | 1606.954 | 0.002 158 40 |
| 9:17:12 | 1606.952 | 489.0577 | 1606.952 | 0.002 158 40 |
| 9:17:22 | 1606.951 | 499.0385 | 1606.951 | 0.002 158 40 |
| 9:17:32 | 1606.949 | 509.0192 | 1606.949 | 0.002 158 40 |
| 9:17:42 | 1606.948 | 519.0000 | 1606.948 | 0.002 158 40 |
| 9:17:52 | 1606.946 | 528.9808 | 1606.946 | 0.002 158 40 |
| 9:18:02 | 1606.945 | 538.9615 | 1606.945 | 0.002 158 40 |
| 9:18:12 | 1606.943 | 548.9423 | 1606.943 | 0.002 158 40 |
| 9:18:22 | 1606.942 | 558.9231 | 1606.942 | 0.002 158 40 |
| 9:18:32 | 1606.970 | 568.9038 | 1606.970 | 0.002 158 40 |
| 9:18:42 | 1607.033 | 578.8846 | 1607.033 | 0.002 158 40 |
| 9:18:52 | 1607.095 | 588.8654 | 1607.095 | 0.002 158 40 |
| 9:19:02 | 1607.159 | 598.8462 | 1607.159 | 0.002 158 40 |
| 9:19:12 | 1607.222 | 608.8269 | 1607.222 | 0.002 158 40 |
| 9:19:22 | 1607.284 | 618.8077 | 1607.284 | 0.002 158 40 |
| 9:19:32 | 1607.347 | 628.7885 | 1607.347 | 0.002 158 40 |
| 9:19:42 | 1607.409 | 638.7692 | 1607.409 | 0.002 158 40 |
| 9:19:52 | 1607.472 | 648.7500 | 1607.472 | 0.002 158 40 |
| 9:20:02 | 1607.534 | 658.7308 | 1607.534 | 0.002 158 40 |
| 9:20:12 | 1607.597 | 668.7115 | 1607.597 | 0.002 158 40 |
| 9:20:22 | 1607.659 | 678.6923 | 1607.659 | 0.002 158 40 |
| 9:20:32 | 1607.722 | 688.6731 | 1607.722 | 0.002 158 40 |
| 9:20:41 | 1607.785 | 698.6538 | 1607.785 | 0.002 158 40 |
| 9:20:51 | 1607.847 | 708.6346 | 1607.847 | 0.002 158 40 |
| 9:21:01 | 1607.912 | 718.6154 | 1607.912 | 0.002 158 40 |
| 9:21:11 | 1607.975 | 728.5962 | 1607.975 | 0.002 158 40 |
| 9:21:21 | 1608.038 | 738.5769 | 1608.038 | 0.002 158 40 |
| 9:21:31 | 1608.100 | 748.5577 | 1608.100 | 0.002 158 40 |
| 9:21:41 | 1608.163 | 758.5385 | 1608.163 | 0.002 158 40 |
| 9:21:51 | 1608.226 | 768.5192 | 1608.226 | 0.002 158 40 |
| 9:22:01 | 1608.289 | 778.5000 | 1608.289 | 0.002 158 40 |
| 9:22:11 | 1608.351 | 788.4808 | 1608.351 | 0.002 158 40 |
| 9:22:21 | 1608.414 | 798.4615 | 1608.414 | 0.002 158 40 |
| 9:22:31 | 1608.477 | 808.4423 | 1608.477 | 0.002 158 40 |
| 9:22:54 | 1608.625 | 818.4231 | 1608.625 | 0.002 158 40 |
| 9:23:04 | 1608.646 | 828.4038 | 1608.646 | 0.002 158 40 |
| 9:23:14 | 1608.646 | 838.3846 | 1608.646 | 0.002 158 40 |
| 9:23:24 | 1608.645 | 848.3654 | 1608.645 | 0.002 158 40 |
| 9:23:34 | 1608.644 | 858.3462 | 1608.644 | 0.002 158 40 |
| 9:23:44 | 1608.643 | 868.3269 | 1608.643 | 0.002 158 40 |
| 9:23:54 | 1608.642 | 878.3077 | 1608.642 | 0.002 158 40 |
| 9:24:04 | 1608.641 | 888.2885 | 1608.641 | 0.002 158 40 |
| 9:24:14 | 1608.640 | 898.2692 | 1608.640 | 0.002 158 40 |
| 9:24:24 | 1608.639 | 908.2500 | 1608.639 | 0.002 158 40 |
| 9:24:34 | 1608.638 | 918.2308 | 1608.638 | 0.002 158 40 |
| 9:24:44 | 1608.638 | 928.2115 | 1608.638 | 0.002 158 40 |
| 9:24:54 | 1608.637 | 938.1923 | 1608.637 | 0.002 158 40 |
| 9:25:04 | 1608.636 | 948.1731 | 1608.636 | 0.002 158 40 |
| 9:25:14 | 1608.635 | 958.1538 | 1608.635 | 0.002 158 40 |
| 9:25:24 | 1608.634 | 968.1346 | 1608.634 | 0.002 158 40 |
| 9:25:34 | 1608.633 | 978.1154 | 1608.633 | 0.002 158 40 |
| 9:25:44 | 1608.632 | 988.0962 | 1608.632 | 0.002 158 40 |
| 9:25:54 | 1608.631 | 998.0769 | 1608.631 | 0.002 158 40 |
| 9:26:04 | 1608.630 | 1008.0577 | 1608.630 | 0.002 158 40 |
| 9:26:14 | 1608.629 | 1018.0385 | 1608.629 | 0.002 158 40 |
| 9:26:24 | 1608.628 | 1028.0192 | 1608.628 | 0.002 158 40 |
| 9:26:34 | 1608.628 | 1038.0000 | 1608.628 | 0.002 158 40 |
| 9:26:44 | 1608.626 | 1047.9808 | 1608.626 | 0.002 158 40 |
| Experiments 3 and 4 | | | | |
| 10:00:48 | 1609.442 | 0.0000 | 1609.436 | 0.003 038 09 |
| 10:00:58 | 1609.440 | 9.9808 | 1609.434 | 0.003 038 09 |
| 10:01:08 | 1609.438 | 19.9615 | 1609.432 | 0.003 038 08 |
| 10:01:18 | 1609.436 | 29.9423 | 1609.430 | 0.003 038 08 |
| 10:01:28 | 1609.434 | 39.9231 | 1609.428 | 0.003 038 07 |
| 10:01:38 | 1609.432 | 49.9038 | 1609.426 | 0.003 038 07 |
| 10:01:48 | 1609.430 | 59.8846 | 1609.424 | 0.003 038 06 |
| 10:01:58 | 1609.428 | 69.8654 | 1609.422 | 0.003 038 06 |
| 10:02:08 | 1609.426 | 79.8462 | 1609.420 | 0.003 038 05 |
| 10:02:18 | 1609.424 | 89.8269 | 1609.418 | 0.003 038 05 |
| 10:02:28 | 1609.422 | 99.8077 | 1609.416 | 0.003 038 05 |
| 10:02:38 | 1609.420 | 109.7885 | 1609.414 | 0.003 038 04 |
| 10:02:48 | 1609.418 | 119.7692 | 1609.412 | 0.003 038 04 |
| 10:02:58 | 1609.416 | 129.7500 | 1609.410 | 0.003 038 03 |
| 10:03:08 | 1609.414 | 139.7308 | 1609.408 | 0.003 038 03 |
| 10:03:18 | 1609.411 | 149.7115 | 1609.405 | 0.003 038 02 |
| 10:03:28 | 1609.409 | 159.6923 | 1609.403 | 0.003 038 02 |
| 10:03:38 | 1609.407 | 169.6731 | 1609.401 | 0.003 038 01 |
| 10:03:48 | 1609.405 | 179.6538 | 1609.399 | 0.003 038 01 |
| 10:03:58 | 1609.403 | 189.6346 | 1609.397 | 0.003 038 01 |
| 10:04:08 | 1609.401 | 199.6154 | 1609.395 | 0.003 038 00 |
| 10:04:18 | 1609.425 | 209.5962 | 1609.419 | 0.003 038 05 |
| 10:04:28 | 1609.489 | 219.5769 | 1609.483 | 0.003 038 19 |
| 10:04:38 | 1609.550 | 229.5577 | 1609.544 | 0.003 038 32 |
| 10:04:48 | 1609.612 | 239.5385 | 1609.606 | 0.003 038 46 |
| 10:04:58 | 1609.673 | 249.5192 | 1609.667 | 0.003 038 59 |
| 10:05:08 | 1609.735 | 259.5000 | 1609.729 | 0.003 038 72 |
| 10:05:18 | 1609.796 | 269.4808 | 1609.790 | 0.003 038 85 |
| 10:05:28 | 1609.858 | 279.4615 | 1609.852 | 0.003 038 99 |
| 10:05:38 | 1609.919 | 289.4423 | 1609.913 | 0.003 039 12 |
| 10:05:48 | 1609.981 | 299.4231 | 1609.975 | 0.003 039 25 |
| 10:05:58 | 1610.042 | 309.4038 | 1610.036 | 0.003 039 38 |
| 10:06:08 | 1610.104 | 319.3846 | 1610.098 | 0.003 039 52 |
| 10:06:18 | 1610.165 | 329.3654 | 1610.159 | 0.003 039 65 |
| 10:06:28 | 1610.229 | 339.3462 | 1610.223 | 0.003 039 79 |
| 10:06:38 | 1610.290 | 349.3269 | 1610.284 | 0.003 039 92 |
| 10:06:48 | 1610.351 | 359.3077 | 1610.345 | 0.003 040 05 |
| 10:06:58 | 1610.413 | 369.2885 | 1610.407 | 0.003 040 18 |
| 10:07:08 | 1610.474 | 379.2692 | 1610.468 | 0.003 040 32 |
| 10:07:18 | 1610.535 | 389.2500 | 1610.529 | 0.003 040 45 |
| 10:07:27 | 1610.597 | 399.2308 | 1610.591 | 0.003 040 58 |
| 10:07:37 | 1610.658 | 409.2115 | 1610.652 | 0.003 040 71 |
| 10:07:47 | 1610.720 | 419.1923 | 1610.714 | 0.003 040 85 |
| 10:07:57 | 1610.781 | 429.1731 | 1610.775 | 0.003 040 98 |
| 10:08:07 | 1610.842 | 439.1538 | 1610.836 | 0.003 041 11 |
| 10:08:17 | 1610.861 | 449.1346 | 1610.855 | 0.003 041 15 |
| 10:08:27 | 1610.859 | 459.1154 | 1610.853 | 0.003 041 15 |
| 10:08:37 | 1610.857 | 469.0962 | 1610.851 | 0.003 041 14 |
| 10:08:47 | 1610.854 | 479.0769 | 1610.848 | 0.003 041 14 |
| 10:08:57 | 1610.852 | 489.0577 | 1610.846 | 0.003 041 13 |
| 10:09:07 | 1610.850 | 499.0385 | 1610.844 | 0.003 041 13 |
| 10:09:17 | 1610.848 | 509.0192 | 1610.842 | 0.003 041 12 |
| 10:09:27 | 1610.846 | 519.0000 | 1610.840 | 0.003 041 12 |
| 10:09:37 | 1610.844 | 528.9808 | 1610.838 | 0.003 041 11 |
| 10:09:47 | 1610.842 | 538.9615 | 1610.836 | 0.003 041 11 |
| 10:09:57 | 1610.840 | 548.9423 | 1610.834 | 0.003 041 11 |
| 10:10:07 | 1610.838 | 558.9231 | 1610.832 | 0.003 041 10 |
| 10:10:17 | 1610.836 | 568.9038 | 1610.830 | 0.003 041 10 |
| 10:10:27 | 1610.834 | 578.8846 | 1610.828 | 0.003 041 09 |
| 10:10:37 | 1610.832 | 588.8654 | 1610.826 | 0.003 041 09 |
| 10:10:47 | 1610.830 | 598.8462 | 1610.824 | 0.003 041 08 |
| 10:10:57 | 1610.828 | 608.8269 | 1610.822 | 0.003 041 08 |
| 10:11:07 | 1610.826 | 618.8077 | 1610.820 | 0.003 041 08 |
| 10:11:17 | 1610.824 | 628.7885 | 1610.818 | 0.003 041 07 |
| 10:11:27 | 1610.822 | 638.7692 | 1610.816 | 0.003 041 07 |
| 10:11:37 | 1610.820 | 648.7500 | 1610.814 | 0.003 041 06 |
| 10:11:47 | 1610.818 | 658.7308 | 1610.812 | 0.003 041 06 |
| 10:11:57 | 1610.837 | 668.7115 | 1610.831 | 0.003 041 10 |
| 10:12:07 | 1610.897 | 678.6923 | 1610.891 | 0.003 041 23 |
| 10:12:17 | 1610.960 | 688.6731 | 1610.954 | 0.003 041 36 |
| 10:12:27 | 1611.021 | 698.6538 | 1611.015 | 0.003 041 50 |
| 10:12:37 | 1611.063 | 708.6346 | 1611.057 | 0.003 041 59 |
| 10:12:47 | 1611.062 | 718.6154 | 1611.056 | 0.003 041 58 |
| 10:12:57 | 1611.061 | 728.5962 | 1611.055 | 0.003 041 58 |
| 10:13:07 | 1611.059 | 738.5769 | 1611.053 | 0.003 041 58 |
| 10:13:17 | 1611.057 | 748.5577 | 1611.051 | 0.003 041 57 |
| 10:13:27 | 1611.055 | 758.5385 | 1611.049 | 0.003 041 57 |
| 10:13:37 | 1611.053 | 768.5192 | 1611.047 | 0.003 041 57 |
| 10:13:47 | 1611.052 | 778.5000 | 1611.046 | 0.003 041 56 |
| 10:13:57 | 1611.050 | 788.4808 | 1611.044 | 0.003 041 56 |
| 10:14:07 | 1611.048 | 798.4615 | 1611.042 | 0.003 041 55 |
| 10:14:17 | 1611.046 | 808.4423 | 1611.040 | 0.003 041 55 |
| 10:14:27 | 1611.045 | 818.4231 | 1611.039 | 0.003 041 55 |
| 10:14:37 | 1611.043 | 828.4038 | 1611.037 | 0.003 041 54 |
| 10:14:47 | 1611.042 | 838.3846 | 1611.036 | 0.003 041 54 |
| 10:14:57 | 1611.040 | 848.3654 | 1611.034 | 0.003 041 54 |
| 10:15:07 | 1611.038 | 858.3462 | 1611.032 | 0.003 041 53 |
| 10:15:17 | 1611.037 | 868.3269 | 1611.031 | 0.003 041 53 |
| 10:15:27 | 1611.035 | 878.3077 | 1611.029 | 0.003 041 53 |
| 10:15:37 | 1611.033 | 888.2885 | 1611.027 | 0.003 041 52 |
| 10:15:47 | 1611.032 | 898.2692 | 1611.026 | 0.003 041 52 |
| 10:15:56 | 1611.030 | 908.2500 | 1611.024 | 0.003 041 52 |
| 10:16:06 | 1611.029 | 918.2308 | 1611.023 | 0.003 041 51 |
| 10:16:16 | 1611.027 | 928.2115 | 1611.021 | 0.003 041 51 |
| Experiments 5 and 6 | | | | |
| 10:20:28 | 1611.591 | 0.0000 | 1611.585 | 0.003 042 73 |
| 10:20:38 | 1611.589 | 9.9808 | 1611.583 | 0.003 042 72 |
| 10:20:48 | 1611.588 | 19.9615 | 1611.582 | 0.003 042 72 |
| 10:20:58 | 1611.587 | 29.9423 | 1611.581 | 0.003 042 72 |
| 10:21:08 | 1611.585 | 39.9231 | 1611.579 | 0.003 042 71 |
| 10:21:18 | 1611.584 | 49.9038 | 1611.578 | 0.003 042 71 |
| 10:21:28 | 1611.583 | 59.8846 | 1611.577 | 0.003 042 71 |
| 10:21:38 | 1611.582 | 69.8654 | 1611.576 | 0.003 042 71 |
| 10:21:48 | 1611.580 | 79.8462 | 1611.574 | 0.003 042 70 |
| 10:21:58 | 1611.579 | 89.8269 | 1611.573 | 0.003 042 70 |
| 10:22:08 | 1611.577 | 99.8077 | 1611.571 | 0.003 042 70 |
| 10:22:18 | 1611.576 | 109.7885 | 1611.570 | 0.003 042 69 |
| 10:22:28 | 1611.574 | 119.7692 | 1611.568 | 0.003 042 69 |
| 10:22:38 | 1611.573 | 129.7500 | 1611.567 | 0.003 042 69 |
| 10:22:48 | 1611.571 | 139.7308 | 1611.565 | 0.003 042 68 |
| 10:22:58 | 1611.570 | 149.7115 | 1611.564 | 0.003 042 68 |
| 10:23:08 | 1611.568 | 159.6923 | 1611.562 | 0.003 042 68 |
| 10:23:18 | 1611.567 | 169.6731 | 1611.561 | 0.003 042 67 |
| 10:23:28 | 1611.565 | 179.6538 | 1611.559 | 0.003 042 67 |
| 10:23:38 | 1611.563 | 189.6346 | 1611.557 | 0.003 042 67 |
| 10:23:48 | 1611.562 | 199.6154 | 1611.556 | 0.003 042 66 |
| 10:23:58 | 1611.585 | 209.5962 | 1611.579 | 0.003 042 71 |
| 10:24:08 | 1611.648 | 219.5769 | 1611.642 | 0.003 042 85 |
| 10:24:18 | 1611.712 | 229.5577 | 1611.706 | 0.003 042 99 |
| 10:24:28 | 1611.778 | 239.5385 | 1611.772 | 0.003 043 13 |
| 10:24:38 | 1611.841 | 249.5192 | 1611.835 | 0.003 043 27 |
| 10:24:48 | 1611.904 | 259.5000 | 1611.898 | 0.003 043 40 |
| 10:24:58 | 1611.968 | 269.4808 | 1611.962 | 0.003 043 54 |
| 10:25:08 | 1612.031 | 279.4615 | 1612.025 | 0.003 043 68 |
| 10:25:18 | 1612.094 | 289.4423 | 1612.088 | 0.003 043 81 |
| 10:25:28 | 1612.157 | 299.4231 | 1612.151 | 0.003 043 95 |
| 10:25:38 | 1612.221 | 309.4038 | 1612.215 | 0.003 044 09 |
| 10:25:48 | 1612.284 | 319.3846 | 1612.278 | 0.003 044 22 |
| 10:25:58 | 1612.347 | 329.3654 | 1612.341 | 0.003 044 36 |
| 10:26:08 | 1612.410 | 339.3462 | 1612.404 | 0.003 044 49 |
| 10:26:18 | 1612.473 | 349.3269 | 1612.467 | 0.003 044 63 |
| 10:26:27 | 1612.539 | 359.3077 | 1612.533 | 0.003 044 77 |
| 10:26:37 | 1612.602 | 369.2885 | 1612.596 | 0.003 044 91 |
| 10:26:47 | 1612.665 | 379.2692 | 1612.659 | 0.003 045 05 |
| 10:26:57 | 1612.728 | 389.2500 | 1612.722 | 0.003 045 18 |
| 10:27:07 | 1612.790 | 399.2308 | 1612.784 | 0.003 045 32 |
| 10:27:17 | 1612.854 | 409.2115 | 1612.848 | 0.003 045 45 |
| 10:27:27 | 1612.916 | 419.1923 | 1612.910 | 0.003 045 59 |
| 10:27:37 | 1612.979 | 429.1731 | 1612.973 | 0.003 045 72 |
| 10:27:47 | 1613.042 | 439.1538 | 1613.036 | 0.003 045 86 |
| 10:27:57 | 1613.087 | 449.1346 | 1613.081 | 0.003 045 96 |
| 10:28:07 | 1613.086 | 459.1154 | 1613.080 | 0.003 045 95 |
| 10:28:17 | 1613.084 | 469.0962 | 1613.078 | 0.003 045 95 |
| 10:28:27 | 1613.081 | 479.0769 | 1613.075 | 0.003 045 94 |
| 10:28:37 | 1613.079 | 489.0577 | 1613.073 | 0.003 045 94 |
| 10:28:47 | 1613.076 | 499.0385 | 1613.070 | 0.003 045 93 |
| 10:28:57 | 1613.074 | 509.0192 | 1613.068 | 0.003 045 93 |
| 10:29:07 | 1613.071 | 519.0000 | 1613.065 | 0.003 045 92 |
| 10:29:17 | 1613.068 | 528.9808 | 1613.062 | 0.003 045 92 |
| 10:29:27 | 1613.066 | 538.9615 | 1613.060 | 0.003 045 91 |
| 10:29:37 | 1613.063 | 548.9423 | 1613.057 | 0.003 045 90 |
| 10:29:47 | 1613.061 | 558.9231 | 1613.055 | 0.003 045 90 |
| 10:29:57 | 1613.058 | 568.9038 | 1613.052 | 0.003 045 89 |
| 10:30:07 | 1613.055 | 578.8846 | 1613.049 | 0.003 045 89 |
| 10:30:17 | 1613.053 | 588.8654 | 1613.047 | 0.003 045 88 |
| 10:30:27 | 1613.050 | 598.8462 | 1613.044 | 0.003 045 88 |
| 10:30:37 | 1613.047 | 608.8269 | 1613.041 | 0.003 045 87 |
| 10:30:47 | 1613.045 | 618.8077 | 1613.039 | 0.003 045 87 |
| 10:30:57 | 1613.042 | 628.7885 | 1613.036 | 0.003 045 86 |
| 10:31:07 | 1613.039 | 638.7692 | 1613.033 | 0.003 045 85 |
| 10:31:17 | 1613.037 | 648.7500 | 1613.031 | 0.003 045 85 |
| 10:31:27 | 1613.034 | 658.7308 | 1613.028 | 0.003 045 84 |
| 10:31:37 | 1613.031 | 668.7115 | 1613.025 | 0.003 045 84 |
| 10:31:47 | 1613.029 | 678.6923 | 1613.023 | 0.003 045 83 |
| 10:31:57 | 1613.039 | 688.6731 | 1613.033 | 0.003 045 85 |
| 10:32:07 | 1613.104 | 698.6538 | 1613.098 | 0.003 045 99 |
| 10:32:17 | 1613.168 | 708.6346 | 1613.162 | 0.003 046 13 |
| 10:32:27 | 1613.229 | 718.6154 | 1613.223 | 0.003 046 26 |
| 10:32:37 | 1613.275 | 728.5962 | 1613.269 | 0.003 046 36 |
| 10:32:47 | 1613.275 | 738.5769 | 1613.269 | 0.003 046 36 |
| 10:32:57 | 1613.272 | 748.5577 | 1613.266 | 0.003 046 36 |
| 10:33:07 | 1613.269 | 758.5385 | 1613.263 | 0.003 046 35 |
| 10:33:17 | 1613.266 | 768.5192 | 1613.260 | 0.003 046 34 |
| 10:33:27 | 1613.263 | 778.5000 | 1613.257 | 0.003 046 34 |
| 10:33:37 | 1613.261 | 788.4808 | 1613.255 | 0.003 046 33 |
| 10:33:47 | 1613.258 | 798.4615 | 1613.252 | 0.003 046 33 |
| 10:33:57 | 1613.255 | 808.4423 | 1613.249 | 0.003 046 32 |
| 10:34:07 | 1613.252 | 818.4231 | 1613.246 | 0.003 046 31 |
| 10:34:17 | 1613.250 | 828.4038 | 1613.244 | 0.003 046 31 |
| 10:34:27 | 1613.247 | 838.3846 | 1613.241 | 0.003 046 30 |
| 10:34:37 | 1613.244 | 848.3654 | 1613.238 | 0.003 046 30 |
| 10:34:47 | 1613.241 | 858.3462 | 1613.235 | 0.003 046 29 |
| 10:34:56 | 1613.239 | 868.3269 | 1613.233 | 0.003 046 29 |
| 10:35:06 | 1613.236 | 878.3077 | 1613.230 | 0.003 046 28 |
| 10:35:16 | 1613.233 | 888.2885 | 1613.227 | 0.003 046 27 |
| 10:35:26 | 1613.230 | 898.2692 | 1613.224 | 0.003 046 27 |
| 10:35:36 | 1613.228 | 908.2500 | 1613.222 | 0.003 046 26 |
| 10:35:46 | 1613.225 | 918.2308 | 1613.219 | 0.003 046 25 |
| 10:35:56 | 1613.222 | 928.2115 | 1613.216 | 0.003 046 25 |
| 10:36:06 | 1613.220 | 938.1923 | 1613.214 | 0.003 046 24 |
| 10:36:16 | 1613.217 | 948.1731 | 1613.211 | 0.003 046 24 |

[e]The first two columns are source data and the last three columns are data after correction or calibration. Shading is (dark purple shading) continuous jetting, (light purple shading) burst jetting, and (no shading) before and after baseline period.

**Table S11.** Number concentrations of single microdroplets.

| Experiment | Deposition number | Particle count | Microdroplet volume (pL) | Volume uncertainty (pL) | Concentration (particles/mL) | Concentration uncertainty | |
|---|---|---|---|---|---|---|---|
| | | | | | | (particles/mL) | (%) |
| 200 nm | 1 | 8 | 47.17 | ± 0.46 | $4.6\times10^9$ | ± $4.5\times10^7$ | ± 1.0 |
| | 2 | 6 | 47.18 | ± 0.46 | $3.7\times10^9$ | ± $3.6\times10^7$ | ± 1.0 |
| | 3 | 11 | 47.15 | ± 0.46 | $3.4\times10^9$ | ± $3.2\times10^7$ | ± 1.0 |
| | 4 | 8 | 47.16 | ± 0.46 | $2.1\times10^9$ | ± $2.1\times10^7$ | ± 1.0 |
| | 5 | 10 | 47.17 | ± 0.45 | $1.4\times10^9$ | ± $1.4\times10^7$ | ± 1.0 |
| | 6 | 7 | 47.16 | ± 0.45 | $1.2\times10^9$ | ± $1.1\times10^7$ | ± 1.0 |
| | 7 | 6 | 47.16 | ± 0.45 | $6.1\times10^8$ | ± $5.9\times10^6$ | ± 1.0 |
| | 8 | 6 | 47.14 | ± 0.45 | $5.1\times10^8$ | ± $4.9\times10^6$ | ± 1.0 |
| | 9 | 10 | 47.15 | ± 0.46 | $5.5\times10^8$ | ± $5.4\times10^6$ | ± 1.0 |
| 500 nm #1 | 1 | 217 | 47.21 | ± 0.45 | $3.2\times10^8$ | ± $3.0\times10^6$ | ± 1.0 |
| | 2 | 174 | 47.14 | ± 0.46 | $1.0\times10^9$ | ± $1.0\times10^7$ | ± 1.0 |
| | 3 | 158 | 47.19 | ± 0.45 | $9.1\times10^8$ | ± $8.8\times10^6$ | ± 1.0 |
| | 4 | 101 | 47.21 | ± 0.46 | $6.1\times10^8$ | ± $5.9\times10^6$ | ± 1.0 |
| | 5 | 67 | 47.20 | ± 0.45 | $3.4\times10^8$ | ± $3.2\times10^6$ | ± 1.0 |
| | 6 | 56 | 47.24 | ± 0.45 | $4.2\times10^8$ | ± $4.1\times10^6$ | ± 1.0 |
| | 7 | 29 | 47.23 | ± 0.45 | $2.8\times10^8$ | ± $2.6\times10^6$ | ± 1.0 |
| | 8 | 24 | 47.20 | ± 0.45 | $3.8\times10^8$ | ± $3.7\times10^6$ | ± 1.0 |
| | 9 | 26 | 47.21 | ± 0.45 | $3.2\times10^8$ | ± $3.0\times10^6$ | ± 1.0 |
| | 10 | 15 | 47.21 | ± 0.45 | $4.0\times10^8$ | ± $3.8\times10^6$ | ± 1.0 |
| | 11 | 49 | 47.15 | ± 0.45 | $4.0\times10^8$ | ± $3.9\times10^6$ | ± 1.0 |
| 500 nm #2 | 1 | 43 | 47.21 | ± 0.45 | $2.8\times10^8$ | ± $2.6\times10^6$ | ± 1.0 |
| | 2 | 29 | 46.95 | ± 0.47 | $1.7\times10^8$ | ± $1.7\times10^6$ | ± 1.0 |
| | 3 | 16 | 46.97 | ± 0.47 | $1.3\times10^8$ | ± $1.3\times10^6$ | ± 1.0 |
| | 4 | 20 | 46.99 | ± 0.46 | $2.3\times10^8$ | ± $2.3\times10^6$ | ± 1.0 |
| | 5 | 13 | 46.95 | ± 0.47 | $1.7\times10^8$ | ± $1.7\times10^6$ | ± 1.0 |
| | 6 | 18 | 46.99 | ± 0.46 | $2.1\times10^8$ | ± $2.1\times10^6$ | ± 1.0 |
| | 7 | 15 | 46.97 | ± 0.47 | $1.5\times10^8$ | ± $1.5\times10^6$ | ± 1.0 |
| | 8 | 19 | 46.99 | ± 0.46 | $1.3\times10^8$ | ± $1.3\times10^6$ | ± 1.0 |
| | 9 | 19 | 46.98 | ± 0.46 | $1.3\times10^8$ | ± $1.3\times10^6$ | ± 1.0 |
| | 10 | 13 | 46.96 | ± 0.47 | $2.1\times10^8$ | ± $2.1\times10^6$ | ± 1.0 |

Uncertainty is the average of the upper and lower 95% coverage intervals.

## Supporting References